\documentclass[11pt]{article}
\usepackage[a4paper, margin=1.2in]{geometry}
\usepackage[english]{babel}
\usepackage[T1]{fontenc}  			
\usepackage{lmodern}
\usepackage[utf8]{inputenc}
\usepackage{microtype}			 	
\usepackage{graphicx}
\usepackage{setspace} 				
\usepackage{tocbibind}				

\usepackage{bm} 					
\usepackage{booktabs}				
\usepackage{array}					%
\usepackage{dcolumn} 				
\usepackage[shortcuts]{extdash} 	

\usepackage[dvipsnames]{xcolor}
\usepackage{pgfplots}
\usepackage{authblk}        
\usepackage{verbatim}       

\usepackage{amsfonts}
\usepackage{amsmath}
\usepackage{amsthm}
\usepackage{amssymb}
\usepackage{mathtools}
\usepackage{commath}
\usepackage{mathrsfs}

\usepackage[caption=false,font=footnotesize]{subfig}
\usepackage{soul}
\usepackage[sort,comma]{natbib}

\usepackage{hyperref}
\hypersetup{colorlinks=true, linkcolor=blue, citecolor=blue}
\usepackage{cleveref}
\usepackage{siunitx}

\usepackage[vlined,lined,ruled]{algorithm2e}
\SetAlgoSkip{bigskip}

\usepackage{notation}

\crefname{algocf}{algorithm}{algorithms}
\Crefname{algocf}{Algorithm}{Algorithms}

\newtheorem{theorem}{Theorem}
\newtheorem{lemma}{Lemma}
\newtheorem{corollary}{Corollary}
\newtheorem{proposition}{Proposition}
\newtheorem{definition}{Definition}

\newtheorem{approximation}{Approximation}

\begin{document}

\title{A Recursive Theory of Variational State Estimation:\\ The Dynamic Programming Approach}

\author{Filip Tronarp}

\date{Centre for Mathematical Sciences, Lund University \\
    \today}

\maketitle

\begin{abstract}
In this article, variational state estimation is examined from the dynamic programming perspective.
This leads to two different value functional recursions depending on whether backward or forward dynamic programming is employed.
The result is a theory of variational state estimation that corresponds to the classical theory of Bayesian state estimation.
More specifically, in the backward method, the value functional corresponds to a likelihood that is upper bounded by the state likelihood
\footnote{This is often referred to as the ``backward filter''.}
from the Bayesian backward recursion.
In the forward method, the value functional corresponds to an unnormalized density that is upper bounded by the unnormalized filtering density.
Both methods can be combined to arrive at a \emph{variational} two-filter formula.
Additionally, it is noted that optimal variational filtering is generally of quadratic time-complexity in the sequence length.
This motivates the notion of sub-optimal variational filtering, which also lower bounds the evidence but is of linear time-complexity.
Another problem is the fact that the value functional recursions are generally intractable.
This is briefly discussed and a simple approximation is suggested that retrieves the filter proposed by \citet{Courts2021}.
The methodology is examined in (i) a jump Gauss--Markov system under a certain factored Markov process approximation, and (ii) in a Gauss--Markov model with log-polynomial likelihoods
under a Gauss--Markov constraint on the variational approximation.
It is demonstrated that the value functional recursions are tractable in both cases.
The resulting estimators are examined in simulation studies and are found to be of adequate quality in comparison to sensible baselines.
\end{abstract}

\section{Introduction}
Consider a partially observed Markov process given by
\begin{subequations}\label{eq:pomp}
\begin{align}
\statevar_0 &\sim \StateDensity_0(\statevar_0) \\
\statevar_t \mid \statevar_{t-1} &\sim \StateDensity_{t\mid t-1}(\statevar_t \mid \statevar_{t-1}) \\
\obsvar_t \mid \statevar_t &\sim \ObsDensity_{t\mid t}(\obsvar_t \mid \statevar_t)
\end{align}
\end{subequations}
on the discrete interval $0 \leq t \leq T$ for some fixed $T$.
Here $\obsvar_{0:T}$ is the observed sequence, $\statevar_{0:T}$ is the (hidden) state sequence, $\StateDensity_0$ is the initial state density,
$\StateDensity_{t\mid t-1}$ are the transition kernels, and $\ObsDensity_{t\mid t}$ is are observation kernels.
The problem of state estimation is to infer the corresponding state sequence $\statevar_{0:T}$.
This kind of estimation problem commonly arise in science and engineering,
such as in tracking and navigation \cite{BarShalom2004,TittertonWeston2004,MarkleyCrassidis2014},
time series analysis \citep{Durbin2012,Hyndman2008},
finance \citep{Lindstrom2018},
robotics \citep{Barfoot2017},
geoscience \citep{Carrassi2018},
data assimilation \citep{Evensen2009},
and neuroscience \citep{Galka2004},
to give a few examples.

From the Bayesian perspective, this entails computing the a posteriori distribution over state paths $\statevar_{0:T}$ given the observed path $\obsvar_{0:T}$.
Formally, the problem is well understood and can be solved by either a backward or forward recursion,
the reader is referred to, for instance, \cite{Ho1964,Kitagawa1987,Cappe2005,Sarkka2023},
a comprehensive account of the linear (Gauss--Markov) theory is given by \citet{Kailath2000}.

However, in general, these recursions involve a series of intractable integrals.
Sequential Monte Carlo methods provide a principled approach to approximating functionals with respect to the posterior over state paths
\citep{Gordon1993,Pitt1999,Arulampalam2002,Crisan2002,Djuric2003,Cappe2007,Kantas2009,Briers2010,Guarniero2017,Naesseth2019}.

A popular and more classical approach to approximate to state estimation is the assumed density method \citep{Maybeck1982},
which ``approximates as you go'' within the Bayesian forward recursion \citep{Bell1993,Julier2000,Lefebvre2002,Wu2006,Tronarp2019b,Corenflos2023,Sarkka2023}.
These approximations can then be refined via iterative schemes \cite{Bell1994,Garcia2015,Garcia2016,Tronarp2018,Garcia2019a}.
However, these approximations can be brittle and are not derived from some principle of optimality,
which make them difficult to analyze.

On the other hand, variational inference allows for control of the class of approximating densities
and is optimal in relative entropy sense \citep{Beal2003,Blei2017}.
Thus the need to \emph{assume} a density is obviated.
While variational inference has long been used in state estimation,
it has mostly been employed as a local approximation within the assumed density framework \citep{Smidl2008,Darling2017,Gultekin2017,Zhao2020,Courts2021}.
Some approaches to \emph{global} variational state estimation has been suggested by \cite{Agamennoni2012,Courts2021},
but these lack the recursive structure akin to Bayesian state estimation theory.
Results in this direction have instead been obtained by \cite{Weber2015},
whom develop a backward recursion for the variational lower bound, see also \cite{Kim2020}.
Further, a forward recursion for the variational lower bound was developed by \cite{Campbell2021},
for which error estimates have been obtained by \citep{Chagneux2024}.

While the approaches of \cite{Weber2015} and \cite{Campbell2021} are indeed recursive,
their relation to classical Bayesian state estimation theory remains opaque.
A central aim of this article is to develop a theory of recursive variational state estimation,
which is completely analogous to Bayesian state estimation theory.
Therefore, before stating the present contribution in full,
a brief review Bayesian state estimation theory is given in \cref{sec:state-estimation-theory},
which also serves to fix some notational conventions used throughout.
The present contribution is then outlined in \cref{sec:contribution}.

\subsection{Bayesian state estimation theory}\label{sec:state-estimation-theory}
There are essentially two ways to construct the a posteriori distribution over state paths,
the forward recursion and the backward recursion \citep{Cappe2005}, see also \citet{Ho1964,Askar1981,Kitagawa1987}.
The result in a \emph{forward} and \emph{reverse-time} Markov factorization of the posterior, respectively.
In order to state the result, introduce the following notation.
The conditional distribution of $\statevar_{t_1:t_2}$ given $\statevar_{t_3:t_4}$ and $\obsvar_{t_5:t_6}$ be denoted by
\begin{equation}
\StateDensity_{t_1:t_2 \mid t_3:t_4}^{t_5:t_6}(\statevar_{t_1:t_2} \mid \statevar_{t_3:t_4}, \obsvar_{t_5:t_6}),
\end{equation}
where in particular $\StateDensity_t^{1:t}$ are the filtering distributions and $\StateDensity_t^{0:T}$ are the smoothing marginals.
Similarly, let the conditional distribution of $\obsvar_{t_1:t_2}$ given $\statevar_{t_3:t_4}$ and $\obsvar_{t_5:t_6}$ be denoted by
\begin{equation}
\ObsDensity_{t_1:t_2 \mid t_3:t_4}^{t_5:t_6}(\obsvar_{t_1:t_2} \mid \statevar_{t_3:t_4}, \obsvar_{t_5:t_6}).
\end{equation}
For example, $\ObsDensity_{1:t}$ is the marginal likelihood associated with $\obsvar_{0:t}$ and $\ObsDensity_{t+1:T\mid t}$ is the likelihood function for $\statevar_t$ given the future observations $\obsvar_{t+1:T}$.
Also, formally define $\ObsDensity_{T+1:T\mid T}(\statevar_T) = 1$.
Furthermore, define the \emph{unnormalized} conditional density of $\statevar_{t_1:t_2}$ given $\statevar_{t_3:t_4}$ and $\obsvar_{t_5:t_6}$ by
\begin{equation}
\UnnormalizedDensity_{t_1:t_2 \mid t_3:t_4}^{t_5:t_6}(\statevar_{t_1:t_2} \mid \statevar_{t_3:t_4}, \obsvar_{t_5:t_6}) =
\ObsDensity_{t_5:t_6 \mid t_1:t_2}(\obsvar_{t_5:t_6} \mid \statevar_{t_1:t_2})
\StateDensity_{t_1:t_2 \mid t_3:t_4}^{t_5:t_6}(\statevar_{t_1:t_2} \mid \statevar_{t_3:t_4}).
\end{equation}
For example, $\UnnormalizedDensity_{t\mid t-1}^t = \ObsDensity_{t\mid t} \StateDensity_{t\mid t-1}$,
the \emph{unnormalized filtering density} satisfy $\UnnormalizedDensity_t^{0:t} = \ObsDensity_{0:t} \StateDensity_t^{0:t}$,
and the \emph{unnormalized smoothing density} satisfy $\UnnormalizedDensity_t^{0:T} = \ObsDensity_{0:T} \StateDensity_t^{0:T}$.
The backward method establishes a recursion over $\ObsDensity_{t+1:T\mid t}$ from which a forward Markov representation of $\StateDensity_{0:T}^{0:T}$ is obtained,
this is \cref{thm:bayesian-backward-recursion}.

\begin{theorem}[Bayesian backward method]
\label{thm:bayesian-backward-recursion}
The likelihood $h_{t+1:T\mid t}$ satisfies the following recursion
\begin{equation}
\ObsDensity_{t:T\mid t-1}(\statevar_{t-1}) = \int \ObsDensity_{t+1:T\mid t}(\statevar_t) \UnnormalizedDensity_{t\mid t-1}^{t}(\statevar_t \mid \statevar_{t-1}) \dif \statevar_t.
\end{equation}
Additionally, the marginal likelihood is given by
\begin{equation}
\ObsDensity_{0:T} = \int \ObsDensity_{1:T\mid 0}(\statevar_0) \UnnormalizedDensity_0^0(\statevar_0) \dif \statevar_0,
\end{equation}
and the state path has the following a posteriori forward Markov representation
\begin{subequations}
\begin{align}
\StateDensity_0^{0:T}(\statevar_0) &= \frac{ \ObsDensity_{1:T\mid 0}(\statevar_0) \UnnormalizedDensity_0^0(\statevar_0)}{\ObsDensity_{0:T}} \\
\StateDensity_{t\mid t-1}^{0:T}(\statevar_t \mid \statevar_{t-1})
&= \frac{\ObsDensity_{t+1:T\mid t}(\statevar_t) \UnnormalizedDensity_{t\mid t-1}^{t}(\statevar_t \mid \statevar_{t-1})}{\ObsDensity_{t:T\mid t-1}(\statevar_{t-1}) }.
\end{align}
\end{subequations}
\end{theorem}

There is a also a corresponding forward method, the so-called filtering recursion.
It is usually stated in terms of the the normalized filtering density, $\StateDensity_t^{0:t}$.
However, in the following, the result is instead stated in terms of a recursion over the unnormalized filtering density, $\UnnormalizedDensity_t^{0:t}$.
The reason is that the unnormalized filtering density is easier to relate to the variational state estimation theory, which is to be developed.
The filtering recursion and its relation to the posterior, $\StateDensity_{0:T}^{0:T}$, is stated in  \cref{thm:bayesian-forward-recursion}.

\begin{theorem}[Bayesian forward method]
\label{thm:bayesian-forward-recursion}
The unnormalized filtering density satisfies the following recursion
\begin{equation}
\UnnormalizedDensity_t^{0:t}(\statevar_t) = \int \UnnormalizedDensity_{t\mid t-1}^{t}(\statevar_t \mid \statevar_{t-1}) \UnnormalizedDensity_{t-1}^{0:t-1}(\statevar_{t-1})  \dif \statevar_{t-1}.
\end{equation}
Additionally, the marginal likelihood is given by
\begin{equation}
\ObsDensity_{0:T} = \int \UnnormalizedDensity_T^{0:T}(\statevar_T) \dif \statevar_T,
\end{equation}
and the state path has the following a posteriori reverse-time Markov representation
\begin{subequations}
\begin{align}
\StateDensity_T^{0:T}(\statevar_T) &= \frac{\UnnormalizedDensity_T^T(\statevar_T)}{\ObsDensity_{0:T}} \\
\StateDensity_{t-1\mid t}^{0:T}(\statevar_{t-1} \mid \statevar_t)
&= \frac{\UnnormalizedDensity_{t\mid t-1}^{t}(\statevar_t \mid \statevar_{t-1}) \UnnormalizedDensity_{t-1}^{0:t-1}(\statevar_{t-1})}{\UnnormalizedDensity_t^{0:t}(\statevar_t)}.
\end{align}
\end{subequations}
\end{theorem}

Lastly, the likelihoods and filtering densities may be combined to obtain the smoothing marginals $\UnnormalizedDensity_t^{0:T}$.
This is the so-called ``two-filter'' formula \citep{Bresler1986}, which is given by \cref{cor:two-filter}.

\begin{corollary}[The two-filter formula]
\label{cor:two-filter}
The unnormalized smoothing density, $\UnnormalizedDensity_t^{0:T}$, satisfy the following relation
\begin{equation}
\UnnormalizedDensity_t^{0:T}(\statevar_t) =  \ObsDensity_{t+1:T\mid t}(\statevar_t) \UnnormalizedDensity_t^{0:t}(\statevar_t).
\end{equation}
\end{corollary}

\subsection{Contribution and outline}\label{sec:contribution}

In variational inference an appropriate class of densities, $\VariationalFamily_{0:T}$, is selected and the posterior is approximated by the best one in the sense of the following lower bound on the marginal likelihood \citep{Beal2003,Blei2017}
\begin{equation*}
\log \ObsDensity_{0:T} \geq \underset{\VariationalDensity_{0:T} \in \VariationalFamily_{0:T}}{\max}\, \int \log\Big\{\frac{\UnnormalizedDensity_{0:T}^{0:T}(\statevar_{0:T})}{\VariationalDensity_{0:T}^{0:T}(\statevar_{0:T})} \Big\}
\VariationalDensity_{0:T}^{0:T}(\statevar_{0:T}) \dif \statevar_{0:T}
\end{equation*}
Recursive approaches have previously been formulated by \citet{Weber2015} and \citet{Campbell2021} from the reinforcement learning perspective.
Here an optimal control approach is taken instead where the time-marginals of $\VariationalDensity_{0:T}$ are states and the associated transition kernels are the controls.
The problem can then formally be solved by both backward and forward dynamic programming \citep{Bellman1966,Larson1966}, which involve recursions over associated value functionals.
\footnote{
It ought to be noted that the present notion of value functionals differs from the notion of value function used by \cite{Weber2015,Campbell2021}.
The relationship between the present contribution and that of \citet{Weber2015,Campbell2021} is further discussed in \cref{sec:related-work}.
}
These methods shall be referred to as the variational backward and forward methods, respectively.
Based on this a theory of recursive variational state estimation is developed in \cref{sec:variational-state-estimation-theory},
which goes beyond the contributions of \citet{Weber2015,Campbell2021}.

More specifically, a variational analogue of the Bayesian state estimation is obtained by establishing:
\begin{itemize}
\item An analogue of \cref{thm:bayesian-backward-recursion} is obtained by the variational backward method.
The associated value functional is completely determined by a likelihood $\BackwardRepresenter_t$, which is upper bounded by $\ObsDensity_{t+1:T\mid t}$.
This is established in \cref{thm:variational-backward-hjb} and \cref{cor:likelihood-lower-bound}.
\item An analogue of \cref{thm:bayesian-forward-recursion} is obtained by the variational forward method.
The associated value functional is completely determined by an unnormalized density $\ForwardRepresenter_t$, which is upper bounded by $\UnnormalizedDensity_t^{0:t}$.
This is established in \cref{thm:variational-forward-hjb} and \cref{cor:filter-lower-bound}.
\item An analogue of \cref{cor:two-filter} is obtained by combining \cref{thm:variational-backward-hjb} and \cref{thm:variational-forward-hjb}.
More specifically, the time-marginals of the variational posterior are relative entropy optimal approximations of the normalization of $\ForwardRepresenter_t\BackwardRepresenter_t$.
This is \cref{cor:variational-two-filter}.
\end{itemize}

The recursions of \cref{thm:variational-backward-hjb} and \cref{thm:bayesian-forward-recursion} are parametrized by the time-marginals of the variational posterior,
which are of course unavailable.
This problem is circumvented by the fixed-point equations of \cref{prop:iterative-backward-smoother,prop:iterative-forward-smoother}.
Furthermore, the notion of variational filtering is discussed in \cref{sec:variational-filtering}.
In particular, the preceeding analysis suggests that the optimal variational filter is in general of time-complexity $O(T^2)$.
This leads to the notion of \emph{sub-optimal} variational filtering, which is of time-compelxity $O(T)$.
Another problem is that the value functional recursions are intractable  in general.
This is briefly discussed in \cref{sec:intractable-value-functionals},
where the fact that $\ForwardRepresenter_t$ is an unnormalized density is used together with variational inference to obtain an approximation.
This procedure is shown to recover the filter proposed by \citet{Courts2021}.

Additionally, two case studies are given where the value functional recursions are indeed tractable.
In \cref{sec:jump-gauss-markov} jump-Gauss Markov systems are studied when the a posterior distribution of the state and jump processes is approximated as independendent
Markov processes and in \cref{sec:logpolynomial} a Gauss--Markov model with log-polynomial likelihoods is examined under a Gauss--Markov constraint.

\section{Variational state estimation theory}\label{sec:variational-state-estimation-theory}
Let $\VariationalFamily_{0:T}$ be a family of distributions over stochastic processes indexed by $0,  \ldots, T$,
and assume that $\UnnormalizedDensity_{0:T}^{0:T}$ absolutely continuous with respect to $\VariationalDensity_{0:T}$ for each $\VariationalDensity_{0:T}$ in $\VariationalFamily_{0:T}$.
Then for any $\VariationalDensity_{0:T} \in \VariationalFamily_{0:T}$, the marginal likelihood, $\ObsDensity_{0:T}$, is given by
\begin{equation}
\ObsDensity_{0:T} = \int \UnnormalizedDensity_{0:T}^{0:T}(\statevar_{0:T}) \dif \statevar_{0:T}
= \int \frac{\UnnormalizedDensity_{0:T}^{0:T}(\statevar_{0:T})}{\VariationalDensity_{0:T}(\statevar_{0:T})}
\VariationalDensity_{0:T}(\statevar_{0:T}) \dif \statevar_{0:T}.
\end{equation}
Furthermore, by Jensen's inequality the logarithm of $\ObsDensity_{0:T}$ is bounded from below by
\begin{equation}
\log \ObsDensity_{0:T} \geq \int \log\Big\{ \frac{\UnnormalizedDensity_{0:T}^{0:T}(\statevar_{0:T})}{\VariationalDensity_{0:T}(\statevar_{0:T})}\Big \}
\VariationalDensity_{0:T}(\statevar_{0:T}) \dif \statevar_{0:T}.
\end{equation}
This lower bound is referred to as \emph{evidence lower bound}, \emph{variational lower bound}, or \emph{free energy} \citep{Beal2003,Blei2017},
which prompts the following definition.
\begin{definition}\label{def:variational-lower-bound}
Given an unnormalized posterior, $\UnnormalizedDensity_{0:T}^{0:T}$, the variational lower bound functional, $\VariationalObjective$, is defined as
\begin{equation}
\VariationalObjective(\UnnormalizedDensity_{0:T}^{0:T}, \VariationalDensity_{0:T}) = \int \log\Big \{ \frac{\UnnormalizedDensity_{0:T}^{0:T}(\statevar_{0:T})}{\VariationalDensity_{0:T}(\statevar_{0:T})}\Big \}
\VariationalDensity(\statevar_{0:T}) \dif \statevar_{0:T}.
\end{equation}
\end{definition}
The idea behind variational inference is then to find the maximizer $\VariationalDensity_{0:T}^{0:T}$ in $\VariationalFamily_{0:T}$ of $\VariationalObjective(\UnnormalizedDensity_{0:T}^{0:T}, \VariationalDensity_{0:T})$ \cite{Beal2003}.
This gives the notion of \emph{variational posterior}, whose formal definition is stated as follows.
\begin{definition}[Variational posterior]
Given a variational lower bound, $\VariationalObjective_{0:T}$, and a family of distributions, $\VariationalFamily_{0:T}$,
the variational posterior is defined as
\begin{equation}
\VariationalDensity_{0:T}^{0:T} =\argmax[\VariationalDensity_{0:T} \in \VariationalFamily_{0:T}]   \VariationalObjective_{0:T}(\UnnormalizedDensity_{0:T}^{0:T}, \VariationalDensity_{0:T}).
\end{equation}
\end{definition}
It is well-known that the variational posterior is the optimal approximation to $\StateDensity_{0:T}^{0:T}$ in relative entropy sense \citep{Beal2003,Blei2017}.

\subsection{Markov Factorization of the variational family}
The problem is then to define a variational family for which the variational posterior is tractable, computationally efficient, and accurate.
From \cref{sec:state-estimation-theory}, it is already known that the posterior, $\StateDensity_{0:T}^{0:T}$, has both a forward and reverse-time Markov representation.
Consequently, it is sufficient to only consider variational families for which each element has a Markov factorization.
Assume that the elements of the variational family, $\VariationalFamily_{0:T}$, admit a forward Markov factorization:
\begin{equation}\label{eq:variational-forward-markov-factorization}
\VariationalDensity_{0:T}(\statevar_{0:T}) = \VariationalDensity_0(\statevar_0) \prod_{t=1}^T \VariationalDensity_{t\mid t-1}(\statevar_t \mid \statevar_{t-1}).
\end{equation}
From $\VariationalFamily_{0:T}$ classes of time-marginals, $\VariationalFamily_t$, and conditional densities, $\VariationalFamily_{t+1:s \mid t}$ for $s > t$, are defined in the obvious way.
For instance $\VariationalDensity_t \in \VariationalFamily_t$ implies that there is a $\VariationalDensity_{0:T} \in \VariationalFamily_{0:T}$ such that
\begin{equation*}
\VariationalDensity_t(\statevar_t) = \int \VariationalDensity_{0:T}(\statevar_{0:T}) \dif \statevar_{0:t-1, t+1T}.
\end{equation*}
In particular this means if $\VariationalDensity_{0:T} \in \VariationalFamily_{0:T}$ then the factors of \eqref{eq:variational-forward-markov-factorization} lies in $\VariationalFamily_0$
and $\VariationalFamily_{t\mid t-1}$, respectively.
Furthermore, \eqref{eq:variational-forward-markov-factorization} also has a reverse-time Markov representation.
Namely, by Bayes' rule
\begin{equation}
\VariationalDensity_{t-1}(\statevar_{t-1}) \VariationalDensity_{t\mid t-1}(\statevar_t \mid \statevar_{t-1}) =
\VariationalDensity_t(\statevar_t) \VariationalDensity_{t-1 \mid t}(\statevar_{t-1} \mid \statevar_t).
\end{equation}
Consequently, any element in $\VariationalFamily_{0:T}$ also admit a \emph{reverse-time} Markov factorization
\begin{equation}\label{eq:variational-reverse-markov-factorization}
\VariationalDensity_{0:T}(\statevar_{0:T}) = \VariationalDensity_T(\statevar_T) \prod_{t=0}^{T-1} \VariationalDensity_{t\mid t+1}(\statevar_t \mid \statevar_{t+1}).
\end{equation}
Define the classes of conditional densities, $\VariationalFamily_{s:t\mid t+1}$ for $s < t$, defined in the analogous way to $\VariationalFamily_{t+1:s \mid t}$,
then the conditional densities in \eqref{eq:variational-reverse-markov-factorization} lie in  $\VariationalFamily_{s:t\mid t+1}$ and $\VariationalDensity_T$ lie in $\VariationalFamily_T$.
The forward and reverse-time Markov factorizations can also be combined like so
\begin{equation}\label{eq:variational-mixed-markov-factorization}
\VariationalDensity_{0:T}(\statevar_{0:T}) =
\VariationalDensity_{0:t-1\mid t}(\statevar_{0:t-1} \mid \statevar_t)
\VariationalDensity_t(\statevar_t)
\VariationalDensity_{t+1:T\mid t}(\statevar_{t+1:T} \mid \statevar_t).
\end{equation}

With these Markov factorizations, efficient algorithms for obtaining the variational posterior can be obtained by means of dynamic programming \citep{Bellman1966}.
The key to this programme is the following decomposition of the variational lower bound
\begin{equation}\label{eq:variational-lower-bound-decomposition}
\VariationalObjective(\UnnormalizedDensity_{0:T}^{0:T}, \VariationalDensity_{0:T}) = \VariationalObjective(\UnnormalizedDensity_{0:t}^{0:t}, \VariationalDensity_{0:t}) +
\int \VariationalObjective(\UnnormalizedDensity_{t+1:T \mid t}^{t+1:T}, \VariationalDensity_{t+1:T\mid t}) \VariationalDensity_t(\statevar_t) \dif \statevar_{t:T}
\end{equation}
where the fact that $\VariationalObjective(\UnnormalizedDensity_{t+1:T \mid t}^{t+1:T}, \VariationalDensity_{t+1:T\mid t})$ is a function in $\statevar_t$ is omitted from the notation.
In \cref{sec:variational-forward-factorization}, the forward Markov factorization~\eqref{eq:variational-forward-markov-factorization} is employed together with backward dynamic programming,
giving a variational anlogue of \cref{thm:bayesian-backward-recursion}.
In \cref{sec:variational-reverse-time-factorization} the reverse-time Markov factorization~\eqref{eq:variational-reverse-markov-factorization} is employed together with forward dynamic programming,
which gives a variational analogue of \cref{thm:bayesian-forward-recursion}.
Lastly, in \cref{sec:variational-two-filter-formula} the mixed forward/reverse-time factorization~\eqref{eq:variational-mixed-markov-factorization} is employed together with both forward and backward dynamic programming to obtain a variational analogue of \cref{cor:two-filter}.

\subsection{Forward Markov factorization of the variational family}\label{sec:variational-forward-factorization}
In this section the factorization~\eqref{eq:variational-forward-markov-factorization} is used in combination with backward dynamic programming,
where the variational inference problem is formally solved by a backward recursion for a particular value function.
This value functional is obtained by maximizing the second term of~\eqref{eq:variational-lower-bound-decomposition} with respect to $\VariationalDensity_{t+1:T\mid t}$,
which gives the notion of \emph{variational backward value functional}.
\begin{definition}[Variational backward value functional]
\label{def:variational-backward-value}
Define the variational backward value functional at time $T$ by $\BackwardValueFunction_T(\VariationalDensity_T) = 0$ and for times $0 \leq t < T$ by
\begin{equation*}
\BackwardValueFunction_t(\VariationalDensity_t) =  \underset{\VariationalDensity_{t+1:T\mid t} \in \VariationalFamily_{t+1:T\mid T}}{\max}\,
\int \VariationalObjective(\UnnormalizedDensity_{t+1:T \mid t}^{t+1:T}, \VariationalDensity_{t+1:T\mid t}) \VariationalDensity_t(\statevar_t) \dif \statevar_{t:T}.
\end{equation*}
\end{definition}
From \cref{def:variational-backward-value}, the evidence lower bound satisfies the following decomposition
\begin{equation*}
\underset{\VariationalDensity_{0:T} \in \VariationalFamily_{0:T} }{\max\,} \VariationalObjective(\UnnormalizedDensity_{0:T}^{0:T}, \VariationalDensity_{0:T}) =
\underset{\VariationalDensity_{0:t} \in \VariationalFamily_{0:t} }{\max\,} \Big[ \VariationalObjective(\UnnormalizedDensity_{0:t \mid t}^{0:t}, \VariationalDensity_{0:t})
+ \BackwardValueFunction_t(\VariationalDensity_t) \Big].
\end{equation*}
In particular, the initial distribution of the variational posterior may be obtained by
\begin{equation}
\VariationalDensity_0^{0:T} =  \arg\,\max_{\VariationalDensity_0 \in \VariationalFamily_0}\,
\Big[ \int \log \Big\{ \frac{\UnnormalizedDensity_0^0(\statevar_0)}{\VariationalDensity_0(\statevar_0)}\Big\} \VariationalDensity_0(\statevar_0) \dif \statevar_0 + \BackwardValueFunction_0(\VariationalDensity_0)   \Big].
\end{equation}
The variational backward value functional satisfies a particular recursion.
More specifically, it is easy to see that
\begin{equation*}
\VariationalObjective(\UnnormalizedDensity_{t:T \mid t-1}^{t:T}, \VariationalDensity_{t:T\mid t-1}) =
 \VariationalObjective(\UnnormalizedDensity_{t \mid t-1}^t, \VariationalDensity_{t\mid t-1})  +
\int \VariationalObjective(\UnnormalizedDensity_{t+1:T \mid t}^{t+1:T}, \VariationalDensity_{t+1:T\mid t}) \VariationalDensity_{t\mid t-1}(\statevar_t \mid \statevar_{t-1}) \dif \statevar_t.
\end{equation*}
Inserting this into \cref{def:variational-backward-value} then gives the variational backward value functional at time $t-1$ according to
\begin{equation*}
\BackwardValueFunction_{t-1}(\VariationalDensity_{t-1}) = \underset{\VariationalDensity_{t\mid t-1} \in \VariationalFamily_{t\mid t-1}}{\max\,}
\Bigg[ \int \VariationalObjective(\UnnormalizedDensity_{t \mid t-1}^t, \VariationalDensity_{t\mid t-1})
\VariationalDensity_{t-1}(\statevar_{t-1}) \dif \statevar_{t-1}  + \BackwardValueFunction_t(\VariationalDensity_t) \Bigg].
\end{equation*}
The boundary constraint $\BackwardValueFunction_T(\VariationalDensity_T) = 0$ of \cref{def:variational-backward-value} ensure that this recursion is valid for $0 < t \leq T$.
It is important to note that $\VariationalDensity_t$ is obtained by marginalizing $\VariationalDensity_{t\mid t-1}$ with respect to $\VariationalDensity_{t-1}$.
Respecting this constraint gives the following result.
\begin{lemma}
\label{lem:variational-backward-hjb}
The variational backward value functional satisfies the following recursion
\begin{equation}
\BackwardValueFunction_{t-1}(\VariationalDensity_{t-1}) = \max_{\VariationalDensity_{t\mid t-1} \in \VariationalFamily_{t \mid t-1}}\,
\Big[  \int \VariationalObjective(\UnnormalizedDensity_{t \mid t-1}^t, \VariationalDensity_{t\mid t-1}) \VariationalDensity_{t-1}(\statevar_{t-1})\dif \statevar_{t-1}
 +  \BackwardValueFunction_t(\VariationalDensity_t)  \Big],
\end{equation}
where the maximization is subject to the following constraint
\begin{equation}\label{eq:variational-forward-marginal-constraint}
\VariationalDensity_t(\statevar_t) = \int \VariationalDensity_{t \mid t-1}(\statevar_t \mid \statevar_{t-1}) \VariationalDensity_{t-1}(\statevar_{t-1}) \dif \statevar_{t-1}.
\end{equation}
\end{lemma}
The recursion of \cref{lem:variational-backward-hjb} is rather opaque but it turns out that the variational backward value functional is in fact a linear functional.
Namely, the next step is to show that $\BackwardValueFunction_t$ admit the following representation
\begin{equation}
\BackwardValueFunction_t(\VariationalDensity_t) = \int \log\{\BackwardRepresenter_t(\statevar_t)\} \VariationalDensity_t(\statevar_t) \dif \statevar_t.
\end{equation}
As the \emph{backward representer} $\BackwardRepresenter_t$ completely determines the backward value functional $\BackwardValueFunction_t$,
they are both referred to as the backward value functional at times.
According to \cref{def:variational-backward-value}, the terminal condition of $\BackwardRepresenter$ must be $\BackwardRepresenter_T(\statevar_T) = 1$.
Furthermore, if $\VariationalDensity_{t \mid t-1} \in \VariationalFamily_{t\mid t-1}$ and $\VariationalDensity_t$ satisfies the constraint \eqref{eq:variational-forward-marginal-constraint},
then $\BackwardValueFunction_t$ is given by
\begin{equation*}
\BackwardValueFunction_t(\VariationalDensity_t) = \int \log \{\BackwardRepresenter_t(\statevar_t)\} \VariationalDensity_{t \mid t-1}(\statevar_t \mid \statevar_{t-1}) \VariationalDensity_{t-1}(\statevar_{t-1}) \dif \statevar_{t-1:t}.
\end{equation*}
Inserting this into the recursion of \cref{lem:variational-backward-hjb},
expanding $\VariationalObjective(\UnnormalizedDensity_{t \mid t-1}^t, \VariationalDensity_{t\mid t-1})$ using \cref{def:variational-lower-bound},
and invoking the principle of induction gives the following result.
\begin{theorem}\label{thm:variational-backward-hjb}
The variational backward value functional satisfies the following recursion
\begin{equation*}
\begin{split}
&\int \log \{\BackwardRepresenter_{t-1}(\statevar_{t-1})\} \VariationalDensity_{t-1}(\statevar_{t-1}) \dif \statevar_{t-1} \\
&=
\underset{\VariationalDensity_{t\mid t-1} \in \VariationalFamily_{t\mid t-1}}{\max}\, \int \Big(\int \log \Big\{\frac{\BackwardRepresenter_t(\statevar_t) \UnnormalizedDensity_{t \mid t-1}^t(\statevar_t \mid \statevar_{t-1})}
{\VariationalDensity_{t \mid t-1}(\statevar_t \mid \statevar_{t-1})}\Big\}
\VariationalDensity_{t \mid t-1}(\statevar_t \mid \statevar_{t-1})  \dif \statevar_t \Big) \VariationalDensity_{t-1}(\statevar_{t-1}) \dif \statevar_{t-1}.
\end{split}
\end{equation*}
Additionally, the initial density of the variational posterior satisfies
\begin{equation*}
\VariationalDensity_0^{0:T} = \argmax[\VariationalDensity_0 \in \VariationalFamily_0]
\int \log \Big\{\frac{\BackwardRepresenter_0(\statevar_0) \UnnormalizedDensity_0^0(\statevar_0)}{\VariationalDensity_0(\statevar_0)}\Big\}
\VariationalDensity_0(\statevar_0) \dif \statevar_0,
\end{equation*}
and the variational lower bound is the corresponding maximia.
\end{theorem}
\Cref{thm:variational-backward-hjb} allows for the recursive computation of a forward Markov factorization of the variational posterior in principle.
However, there is a practical problem that the recursion for the representer $\BackwardRepresenter_t$ depends on the time marginals $\VariationalDensity_0, \ldots, \VariationalDensity_{T-1}$,
which in fact need to be the time-marginals of the variational posterior.
The solution is to alternative between computing value functionals and transition kernels and time-marginals, which lead to the fixed-point equations given by \cref{prop:iterative-backward-smoother}.
In practice, the fixed-point iteration needs to be initialized.
It appears difficult to construct an intelligent initializer based on the backward method
but the forward method to be discussed in \cref{sec:variational-reverse-time-factorization} can be initialized based on sub-optimal variational filtering to be discussed in \cref{sec:variational-filtering}.

\begin{proposition}[Backward variational fixed-point equations]\label{prop:iterative-backward-smoother}
The backward value functional and the variational posterior satisfy the following fixed-point equations for $1 \leq t \leq T$
\begin{subequations}
\begin{align}
\VariationalDensity_{t\mid t-1}^{0:T} &=
\argmax[\VariationalDensity_{t\mid t-1} \in \VariationalFamily_{t\mid t-1}] \int \Big(\int \log \Big\{\frac{\BackwardRepresenter_t \UnnormalizedDensity_{t \mid t-1}^t}
{\VariationalDensity_{t \mid t-1}}\Big\}
\VariationalDensity_{t \mid t-1} \dif \statevar_t \Big) \VariationalDensity_{t-1}^{0:T} \dif \statevar_{t-1} \\
\log \BackwardRepresenter_{t-1} &= \int \log \Big\{\frac{\BackwardRepresenter_t \UnnormalizedDensity_{t \mid t-1}^t}
{\VariationalDensity_{t \mid t-1}^{0:T}}\Big\}
\VariationalDensity_{t \mid t-1}^{0:T}  \dif \statevar_t \\
\VariationalDensity_0^{0:T} &=\argmax[\VariationalDensity_0 \in \VariationalFamily_0]
\int \log \Big\{\frac{\BackwardRepresenter_0 \UnnormalizedDensity_0^0}{\VariationalDensity_0}\Big\}
\VariationalDensity_0 \dif \statevar_0 \\
\VariationalDensity_t^{0:T} &=\int \VariationalDensity_{t\mid t-1}^{0:T} \VariationalDensity_{t-1}^{0:T} \dif \statevar_{t-1}.
\end{align}
\end{subequations}
\end{proposition}

\subsubsection{The backward representer as a likelihood function}
\Cref{thm:variational-backward-hjb} suggest that the backward representers, $\BackwardRepresenter_t$,
serve a similar purpose in the variational state estimation as the likelihood $\ObsDensity_{t+1:T\mid t}$ serve in Bayesian state estimation (c.f. \cref{thm:bayesian-backward-recursion}).
In fact, the former is a lower bound to the latter since by Jensen's inequality
\begin{equation*}
\begin{split}
\log\BackwardRepresenter_{t-1}(\statevar_{t-1}) &=
\int \log \Big\{ \frac{\BackwardRepresenter_t(\statevar_t)\UnnormalizedDensity_{t \mid t-1}^t(\statevar_t \mid \statevar_{t-1})}
{\VariationalDensity_{t \mid t-1}^{0:T}(\statevar_t \mid \statevar_{t-1})}\Big\}
\VariationalDensity_{t \mid t-1}^{0:T}(\statevar_t \mid \statevar_{t-1}) \dif \statevar_t  \\
&\leq \log \int \BackwardRepresenter_t(\statevar_t)\UnnormalizedDensity_{t \mid t-1}^t(\statevar_t \mid \statevar_{t-1}) \dif \statevar_t.
\end{split}
\end{equation*}
Therefore, $\BackwardRepresenter$, satisfies the following difference inequality
\begin{equation}
\BackwardRepresenter_{t-1}(\statevar_{t-1}) \leq \int \BackwardRepresenter_t(\statevar_t)\UnnormalizedDensity_{t \mid t-1}^t(\statevar_t \mid \statevar_{t-1}) \dif \statevar_t,
\end{equation}
but $\ObsDensity_{t+1:T\mid t}$ satisfies this relation with strict equality.
Thus if $\BackwardRepresenter_t(\statevar_t) \leq \ObsDensity_{t+1:T\mid t}(\statevar_t)$ then the same inequality hold for $t-1$ and the inequality hold for the base case, $t = T$ by definition,
hence \cref{cor:likelihood-lower-bound} is established by induction.

\begin{corollary}\label{cor:likelihood-lower-bound}
The likelihood $\ObsDensity_{t+1:T\mid t}$ is lower bounded by the backward representer $\BackwardRepresenter_t$:
\begin{equation*}
\ObsDensity_{t+1:T\mid t}(\statevar_t) \geq \BackwardRepresenter_t(\statevar_t).
\end{equation*}
\end{corollary}

\subsection{Reverse-time Markov factorization of the variational family}\label{sec:variational-reverse-time-factorization}
In this section the factorization~\eqref{eq:variational-reverse-markov-factorization} is used in combination with forward dynamic programming,
where the variational inference problem is formally solved by a forward recursion for a value functional to be determined.
Note that the first term of~\eqref{eq:variational-lower-bound-decomposition} decomposes as
\begin{equation}\label{eq:variational-lower-bound-decomposition-2}
\VariationalObjective(\UnnormalizedDensity_{0:t}^{0:t}, \VariationalDensity_{0:t}) =
\int \VariationalObjective(\UnnormalizedDensity_{0:t}^{0:t}, \VariationalDensity_{0:t-1\mid t}) \VariationalDensity_t(\statevar_t) \dif \statevar_t
- \int \log\{\VariationalDensity_t(\statevar_t)\} \VariationalDensity_t(\statevar_t) \dif \statevar_t
\end{equation}
The \emph{forward} value functional is then obtained by maximizing the first term of~\eqref{eq:variational-lower-bound-decomposition-2} with respect to $\VariationalDensity_{0:t-1\mid t}$.
\begin{definition}[Variational forward value functional]
\label{def:variational-forward-value}
Define the variational forward value functional for $0 < t \leq T$ by
\begin{equation}
\ForwardValueFunction_t(\VariationalDensity_t) =
\max_{\VariationalDensity_{0:t-1 \mid t} \in \VariationalFamily_{0:t-1\mid t}}
\int \VariationalObjective(\UnnormalizedDensity_{0:t}^{0:t}, \VariationalDensity_{0:t-1\mid t}) \VariationalDensity_t(\statevar_t) \dif \statevar_t
\end{equation}
and define the initial condition, $\ForwardValueFunction_0$, by
\begin{equation}
\ForwardValueFunction_0(\VariationalDensity_0) = \int \log \{\UnnormalizedDensity_0^0(\statevar_0)\} \VariationalDensity_0(\statevar_0) \dif \statevar_0.
\end{equation}
\end{definition}
Just as the variational backward value functional satisfies a backward recursion, the variational forward value functional satisfies a forward recursion.
Namely, the following identity holds for $1 \leq t$
\begin{equation*}
\VariationalObjective(\UnnormalizedDensity_{0:t}^{0:t}, \VariationalDensity_{0:t-1\mid t}) =
\int \VariationalObjective(\UnnormalizedDensity_{0:t-1}^{0:t-1}, \VariationalDensity_{0:t-2\mid t-1}) \VariationalDensity_{t-1\mid t}(\statevar_{t-1} \mid \statevar_t) \dif \statevar_{t-1}
+ \VariationalObjective(\UnnormalizedDensity_{t\mid t-1}^t, \VariationalDensity_{t-1\mid t})
\end{equation*}
where $\VariationalObjective(\UnnormalizedDensity_{0}^{0}, \VariationalDensity_{0:-1\mid 0}) = \log \{\UnnormalizedDensity_0^0(\statevar_0)\}$ by convention.
Inserting this into \cref{def:variational-forward-value} then gives variational forward value functional at time $t$ according to
\begin{equation*}
\ForwardValueFunction_t(\VariationalDensity_t) = \max_{\VariationalDensity_{t-1 \mid t} \in \VariationalFamily_{t-1\mid t}}
\Big[
\ForwardValueFunction_{t-1}(\VariationalDensity_{t-1})
+ \int \VariationalObjective(\UnnormalizedDensity_{t\mid t-1}^t, \VariationalDensity_{t-1\mid t})
\VariationalDensity_t(\statevar_t) \dif \statevar_t
\Big].
\end{equation*}
Here $\VariationalDensity_{t-1}$ is obtained by marginalizing $\VariationalDensity_{t-1\mid t}$ with respect to $\VariationalDensity_t$.
Accounting for this constraint gives the following result.
\begin{lemma}
\label{lem:variational-forward-hjb}
The variational forward value functional satisfies the following recursion
\begin{equation}
\ForwardValueFunction_t(\VariationalDensity_t) = \max_{\VariationalDensity_{t-1 \mid t} \in \VariationalFamily_{t-1\mid t}}
\Big[
\ForwardValueFunction_{t-1}(\VariationalDensity_{t-1}) + \int \VariationalObjective(\UnnormalizedDensity_{t\mid t-1}^t, \VariationalDensity_{t-1\mid t})
\VariationalDensity_t(\statevar_t) \dif \statevar_t
\Big],
\end{equation}
where the maximization is subject to the constraint
\begin{equation}\label{eq:variational-backward-marginal-constraint}
\VariationalDensity_{t-1}(\statevar_{t-1}) = \int \VariationalDensity_{t-1\mid t}(\statevar_{t-1} \mid \statevar_t) \VariationalDensity_t(\statevar_t) \dif \statevar_t.
\end{equation}
\end{lemma}
Similarly to the variational backward value functional, the variational forward value functional is a linear functional.
Namely, make the following ansatz
\begin{equation}
\ForwardValueFunction_t(\VariationalDensity_t) = \int \log \{\ForwardRepresenter_t(\statevar_t)\} \VariationalDensity_t(\statevar_t) \dif \statevar_t, \quad 0 \leq t \leq T.
\end{equation}
Then by \cref{def:variational-forward-value}, the initial condition of $\ForwardRepresenter$ is given by $\ForwardRepresenter_0 = \UnnormalizedDensity_0^0$.
Additionally, if $\VariationalDensity_{t-1\mid t} \in \VariationalFamily_{t-1\mid t}$ and $\VariationalDensity_{t-1}$ satisfies the constraint~\eqref{eq:variational-backward-marginal-constraint},
then the forward value functional is at time $t$ given by
\begin{equation}
\ForwardValueFunction_{t-1}(\VariationalDensity_{t-1}) = \int \log \{\ForwardRepresenter_{t-1}(\statevar_{t-1})\} \VariationalDensity_{t-1\mid t}(\statevar_{t-1} \mid \statevar_t) \VariationalDensity_t(\statevar_t) \dif \statevar_{t-1:t}
\end{equation}
Inserting this into the recursion of \cref{lem:variational-forward-hjb} gives the following result by induction.
\begin{theorem}
\label{thm:variational-forward-hjb}
The variational forward value functional satisfies the following recursion
\begin{equation}
\begin{split}
&\int \log \{\ForwardRepresenter_t(\statevar_t)\} \VariationalDensity_t(\statevar_t) \dif \statevar_t \\
&=  \underset{\VariationalDensity_{t-1\mid t} \in \VariationalFamily_{t-1\mid t}}{\max}\,
\int \Big(\int \log\Big\{ \frac{\UnnormalizedDensity_{t\mid t-1}^t(\statevar_t\mid \statevar_{t-1}) \ForwardRepresenter_{t-1}(\statevar_{t-1})}{\VariationalDensity_{t-1\mid t}(\statevar_{t-1} \mid \statevar_t)}
\Big\} \VariationalDensity_{t-1\mid t}(\statevar_{t-1} \mid \statevar_t) \dif \statevar_{t-1}  \Big)
\VariationalDensity_t(\statevar_t) \dif \statevar_t.
\end{split}
\end{equation}
Additionally, the terminal density of the variational posterior satisfies
\begin{equation}
\VariationalDensity_T^{0:T} = \argmax[\VariationalDensity_T \in \VariationalFamily_T] \int  \log \Big\{ \frac{\ForwardRepresenter_T(\statevar_T)}{\VariationalDensity_T(\statevar_T)}\Big\}
\VariationalDensity_T(\statevar_T) \dif \statevar_T,
\end{equation}
and the lower bound is given by the corresponding maxima.
\end{theorem}

\Cref{thm:variational-forward-hjb} does in principle allow for computing a reverse-time Markov factorization of the variational posterior via a \emph{forward} recursion.
However, just as for the backward recursion the value functional recursion relies on the time-marginals $\VariationalDensity_1, \ldots, \VariationalDensity_T$,
which need to be time-marginals of the variational posterior.
Just as for the backward dynamic programming approach a fixed-point iteration can be developed, see \cref{prop:iterative-forward-smoother}.
This algorithm can be initialized by a variational filter as discussed in \cref{sec:variational-filtering}.

\begin{proposition}[Forward variational fixed-point equations]\label{prop:iterative-forward-smoother}
The forward value functional and the variational posterior satisfy the following fixed-point equations for $1 \leq t \leq T$
\begin{subequations}
\begin{align}
\ForwardRepresenter_0 &= \UnnormalizedDensity_0^0 \\
\VariationalDensity_{t\mid t-1}^{0:T} &= \argmax[\VariationalDensity_{t-1\mid t} \in \VariationalFamily_{t-1\mid t}]
\int \Big(\int \log \Big\{\frac{\UnnormalizedDensity_{t \mid t-1}^t \ForwardRepresenter_{t-1}}
{\VariationalDensity_{t-1 \mid t}}\Big\}
\VariationalDensity_{t-1 \mid t} \dif \statevar_{t-1} \Big) \VariationalDensity_t^{0:T} \dif \statevar_t \\
\log \ForwardRepresenter_t &= \int \log \Big\{\frac{ \UnnormalizedDensity_{t \mid t-1}^t \ForwardRepresenter_{t-1}}
{\VariationalDensity_{t-1 \mid t}^{0:T}}\Big\}
\VariationalDensity_{t-1 \mid t}^{0:T}  \dif \statevar_{t-1} \\
\VariationalDensity_T^{0:T} &= \argmax[\VariationalDensity_T \in \VariationalFamily_T]
\int \log \Big\{\frac{\ForwardRepresenter_T}{\VariationalDensity_T}\Big\}
\VariationalDensity_T \dif \statevar_T \\
\VariationalDensity_{t-1}^{0:T} &= \int \VariationalDensity_{t-1\mid t}^{0:T} \VariationalDensity_t^{0:T} \dif \statevar_t.
\end{align}
\end{subequations}
\end{proposition}

\subsubsection{The forward representer as an unnormalized density}
The forward representer, $\ForwardRepresenter_t$ is the variational analogues to the unnormalized filtering density, $\UnnormalizedDensity_t^{0:t}$.
Similarly, to the backward representer they serve as a lower bound to its analogues.
Namely, by Jensen's inequality
\begin{equation}
\begin{split}
\log\ForwardRepresenter_t(\statevar_t) &=
\int \log \Big\{\frac{\UnnormalizedDensity_{t \mid t-1}^t(\statevar_t \mid \statevar_{t-1}) \ForwardRepresenter_{t-1}(\statevar_{t-1})}
{\VariationalDensity_{t-1 \mid t}^{0:T}(\statevar_{t-1} \mid \statevar_t)}\Big\}
 \VariationalDensity_{t-1 \mid t}^{0:T}(\statevar_{t-1} \mid \statevar_t) \dif \statevar_{t-1} \\
&\leq \log\int \UnnormalizedDensity_{t \mid t-1}^t(\statevar_t \mid \statevar_{t-1}) \ForwardRepresenter_{t-1}(\statevar_{t-1})
\dif \statevar_{t-1}.
\end{split}
\end{equation}
Therefore, forward representer satisfy the following difference inequality
\begin{equation}
\ForwardRepresenter_t(\statevar_t) \leq \int \UnnormalizedDensity_{t \mid t-1}^t(\statevar_t \mid \statevar_{t-1}) \ForwardRepresenter_{t-1}(\statevar_{t-1})
\dif \statevar_{t-1}.
\end{equation}
but $\UnnormalizedDensity_t^{0:t}$ satisfy this relation with strict equality.
Hence by  similar induction argument as for \cref{cor:likelihood-lower-bound}, the following bound can be established.
\begin{corollary}
\label{cor:filter-lower-bound}
The unnormalized filtering distribution $\UnnormalizedDensity_t^{0:t}$ and the forward value functional satisfy the following bound
\begin{equation}
\UnnormalizedDensity_t^{0:t}(\statevar_t) \geq \ForwardRepresenter_t(\statevar_t),
\end{equation}
hence, $\ForwardRepresenter_t$ is the density of a positive finite measure.
\end{corollary}

\subsubsection{The value functional recursion in predictor-corrector form}
It ought to be noted that the forward value functional recursion can be written in so-called predictor-corrector form.
Inserting the identity $\UnnormalizedDensity_{t\mid t-1}^t = \ObsDensity_{t\mid t} \StateDensity_{t\mid t-1}$ into \cref{thm:variational-forward-hjb} gives
\begin{equation*}
\begin{split}
&\int \log \{\ForwardRepresenter_t(\statevar_t)\} \VariationalDensity_t(\statevar_t) \dif \statevar_t = \int \log\{\ObsDensity_{t\mid t}(\statevar_t)\} \VariationalDensity_t(\statevar_t) \dif \statevar_t\\
&+  \underset{\VariationalDensity_{t-1\mid t} \in \VariationalFamily_{t-1\mid t}}{\max}\,
\int \Big(\int \log\Big\{ \frac{\StateDensity_{t\mid t-1}^t(\statevar_t\mid \statevar_{t-1}) \ForwardRepresenter_{t-1}(\statevar_{t-1})}{\VariationalDensity_{t-1\mid t}(\statevar_{t-1} \mid \statevar_t)}
\Big\} \VariationalDensity_{t-1\mid t}(\statevar_{t-1} \mid \statevar_t) \dif \statevar_{t-1}  \Big)
\VariationalDensity_t(\statevar_t) \dif \statevar_t.
\end{split}
\end{equation*}
From which the following result is immediate.
\begin{proposition}\label{prop:forward-value-functional-predictor-corrector}
The forward value functional satisfy the following predictor-corrector recursion:
\begin{subequations}
\begin{align*}
\begin{split}
&\int \log \{\ForwardRepresenter_t^{\mathsf{p}}(\statevar_t)\} \VariationalDensity_t(\statevar_t) \dif \statevar_t\\
&= \underset{\VariationalDensity_{t-1\mid t} \in \VariationalFamily_{t-1\mid t}}{\max}\,
\int \Big(\int \log\Big\{ \frac{\StateDensity_{t\mid t-1}(\statevar_t\mid \statevar_{t-1}) \ForwardRepresenter_{t-1}(\statevar_{t-1})}{\VariationalDensity_{t-1\mid t}(\statevar_{t-1} \mid \statevar_t)}
\Big\} \VariationalDensity_{t-1\mid t}(\statevar_{t-1} \mid \statevar_t) \dif \statevar_{t-1}  \Big)
\VariationalDensity_t(\statevar_t) \dif \statevar_t
\end{split}\\
&\log \ForwardRepresenter_t(\statevar_t) = \log \ForwardRepresenter_t^{\mathsf{p}}(\statevar_t) + \log \ObsDensity_{t\mid t}(\statevar_t).
\end{align*}
\end{subequations}
\end{proposition}

Additionally, a result similar to \cref{cor:filter-lower-bound} for the predictive forward value functional.
That is, from Jensen's inequality and \cref{cor:filter-lower-bound}, the following bound hold
\begin{equation}
\begin{split}
\log \ForwardRepresenter_t^{\mathsf{p}}(\statevar_t)
&\leq \log\int \StateDensity_{t\mid t-1}(\statevar_t \mid \statevar_{t-1})  \ForwardRepresenter_{t-1}(\statevar_{t-1}) \dif \statevar_{t-1} \\
&\leq \log \int \StateDensity_{t\mid t-1}(\statevar_t \mid \statevar_{t-1})  \UnnormalizedDensity_{t-1}^{0:t-1}(\statevar_{t-1})  \dif \statevar_{t-1}
=  \log \UnnormalizedDensity_t^{0:t-1}(\statevar_t).
\end{split}
\end{equation}

\begin{corollary}
\label{cor:predictor-lower-bound}
The unnormalized predictive distribution $\UnnormalizedDensity_t^{0:t-1}$ and the predictive forward value functional satisfy the following bound
\begin{equation}
\ForwardRepresenter_t^{\mathsf{p}}(\statevar_t) \leq \UnnormalizedDensity_t^{0:t-1}(\statevar_t).
\end{equation}
\end{corollary}

\subsection{A Variational two-filter formula}\label{sec:variational-two-filter-formula}
In order to complete the theory of variational state estimation,
a relation corresponding to the two-filter formula of \cref{cor:two-filter} need to be established.
By combining~\eqref{eq:variational-lower-bound-decomposition} and~\eqref{eq:variational-lower-bound-decomposition-2},
the variational lower bound is obtained as
\begin{equation}
\begin{split}
\VariationalObjective_{0:T}(\UnnormalizedDensity_{0:T}^{0:T}, \VariationalDensity_{0:T}) &=
\int \VariationalObjective(\UnnormalizedDensity_{0:t}^{0:t}, \VariationalDensity_{0:t-1\mid t}) \VariationalDensity_t(\statevar_t) \dif \statevar_t
+ \int \VariationalObjective(\UnnormalizedDensity_{t+1:T \mid t}^{t+1:T}, \VariationalDensity_{t+1:T\mid t}) \VariationalDensity_t(\statevar_t) \dif \statevar_{t:T} \\
&\quad - \int \log\{\VariationalDensity_t(\statevar_t)\} \VariationalDensity_t(\statevar_t) \dif \statevar_t.
\end{split}
\end{equation}
Maximizing first with respect to both $\VariationalDensity_{0:t-1\mid t}$ and $\VariationalDensity_{t+1:T \mid t}$ and then maximizing with respect to $\VariationalDensity_t$
gives
\begin{equation}
\VariationalDensity_t^{0:T} =
\argmax[\VariationalDensity_t \in \VariationalFamily_t] \Big[\ForwardValueFunction_t(\VariationalDensity_t) - \int \log\{\VariationalDensity_t(\statevar_t)\} \VariationalDensity_t(\statevar_t) \dif \statevar_t  + \BackwardValueFunction_t(\VariationalDensity_t)  \Big].
\end{equation}
Now writing the value functionals in terms of $\ForwardRepresenter_t$ and $\BackwardRepresenter_t$, respectively, gives the variational two-filter formula.
\begin{corollary}[Variational two-filter smoother]
\label{cor:variational-two-filter}
The variational smoothing density at time $t$ is the maximizer of
\begin{equation}
\VariationalDensity_t^{0:T} = \underset{\VariationalDensity_t \in \VariationalFamily_t}{\arg\,\max\,}
\int \log \Big\{\frac{\ForwardRepresenter_t(\statevar_t)\BackwardRepresenter_t(\statevar_t)}{\VariationalDensity_t(\statevar_t)}\Big\} \VariationalDensity_t(\statevar_t) \dif \statevar_t.
\end{equation}
\end{corollary}

\section{Optimal and sub-optimal variational filtering}\label{sec:variational-filtering}

In many applications it is interesting to compute an approximation of the filtering density, $\StateDensity_t^{0:t}$, on-line.
Note that the dependence of the value functionals with respect to the observation sequence has hitherto been supressed from the notation.
However, in order to appropriately discuss the possibility of variational filtering it needs to be there.
That is, let $\ForwardRepresenter_t^{0:s}$ denote the forward value functional with respect to the observations $\obsvar_\tau, 0 \leq \tau \leq s$.
Now in view of \cref{thm:variational-forward-hjb} and \cref{cor:filter-lower-bound}, the variational filtering densities may be defined as follows.
\begin{definition}[Optimal variational filtering]
\label{def:optimal-variational-filtering}
The variational filtering densities, $\VariationalDensity_t^{0:t}$, with respect to the family $\VariationalFamily_{0:T}$, are given by
\begin{equation}\label{eq:def-variational-filter}
\VariationalDensity_t^{0:t} = \argmax \int \log \Big\{\frac{\ForwardRepresenter_t^{0:t}(\statevar_t)}{\VariationalDensity_t(\statevar_t)}\Big\} \VariationalDensity_t(\statevar_t) \dif \statevar_t.
\end{equation}
\end{definition}

The problem is now immediately clear, \cref{thm:variational-forward-hjb} gives a recursion in $t$ of $\ForwardRepresenter_t^{0:T}$.
However, in order to obtain a recursive variational filter what is needed is a recursion of $\ForwardRepresenter_t^{0:t}$ in $t$.
The present analysis suggests that this is in general impossible without re-processing previous observations.
This indicates that the optimal filter given by \cref{def:optimal-variational-filtering} is of time-complexity $O(T^2)$ (c.f \cref{prop:iterative-forward-smoother}).

However, there is a special case when the optimal variational filter is of time-complexity $O(T)$.
Define the universally optimal predictive value functional, $\ForwardRepresenter_t^\dagger$, and the universally optimal reverse-time transition kernel, $\VariationalDensity_{t-1\mid t}^\dagger$
with respect to $\ForwardRepresenter_{t-1}^{0:t-1}$, by Bayes' rule
\begin{equation}
\ForwardRepresenter_t^\dagger \VariationalDensity_{t-1\mid t}^\dagger =
\StateDensity_{t\mid t-1} \ForwardRepresenter_{t-1}.
\end{equation}
Inserting this into the recursion of \cref{thm:variational-forward-hjb} then gives
\begin{equation*}
\begin{split}
&\int \log \{\ForwardRepresenter_t^{0:t}(\statevar_t)\} \VariationalDensity_t(\statevar_t) \dif \statevar_t =
\int \log\{\ObsDensity_{t\mid t}(\statevar_t) \ForwardRepresenter_t^\dagger(\statevar_t) \} \VariationalDensity_t(\statevar_t) \dif \statevar_t \\
&+  \underset{\VariationalDensity_{t-1\mid t} \in \VariationalFamily_{t-1\mid t}}{\max}\,
\int \Big(\int \log\Big\{ \frac{\VariationalDensity_{t-1\mid t}^\dagger(\statevar_{t-1} \mid \statevar_t)}{\VariationalDensity_{t-1\mid t}(\statevar_{t-1} \mid \statevar_t)}
\Big\} \VariationalDensity_{t-1\mid t}(\statevar_{t-1} \mid \statevar_t) \dif \statevar_{t-1}  \Big)
\VariationalDensity_t(\statevar_t) \dif \statevar_t.
\end{split}
\end{equation*}
If $\VariationalDensity_{t-1\mid t}^\dagger \in \VariationalFamily_{t-1\mid t}$, then the last term on the right-hand side disappears and the following result is evident.
\begin{proposition}[Recursive optimal variational filtering]
\label{prop:optimal-variational-filtering}
Assume that $\VariationalDensity_{t-1\mid t}^\dagger \in \VariationalFamily_{t-1\mid t}$, then the forward value functional satisfies the following recursion for $1 \leq t \leq T$
\begin{subequations}
\begin{align}
\ForwardRepresenter_t^\dagger(\statevar_t) &= \int \StateDensity_{t\mid t-1}(\statevar_t \mid \statevar_{t-1}) \ForwardRepresenter_{t-1}^{0:t-1}(\statevar_{t-1}) \dif \statevar_{t-1}\\
\log \ForwardRepresenter_t^{0:t}(\statevar_t) &= \log\{\ObsDensity_{t\mid t}(\statevar_t) \ForwardRepresenter_t^\dagger(\statevar_t) \}.
\end{align}
\end{subequations}
Additionally, $\ForwardRepresenter_t^{0:t} = \ForwardRepresenter_t^{0:T}$.
\end{proposition}

\Cref{prop:optimal-variational-filtering} is of course just a disguised version of the Bayesian filtering recursion of \cref{thm:bayesian-forward-recursion}.
Therefore, it appears that when $\VariationalFamily_{0:t-1\mid t}$ is not rich enough to contain $\StateDensity_{0:t-1\mid t}$,
then some notion of sub-optimal variational filtering is required.
One such candidate is specified in \cref{def:sub-optimal-variational-filter-1}.

\begin{definition}[Sub-optimal variational filter I]\label{def:sub-optimal-variational-filter-1}
The variational filter of type I satisfies the following equations for $1 \leq t$
\begin{subequations}
\begin{align}
\ForwardRepresenter_0^0 &= \UnnormalizedDensity_0^0 \\
\VariationalDensity_0^0 &= \argmax[\VariationalDensity_0 \in \VariationalFamily_0] \int \log \Big\{ \frac{\ForwardRepresenter_0^0}{\VariationalDensity_0} \Big\}
\VariationalDensity_0\dif \statevar_0 \\
\VariationalDensity_t^{0:t},\, \VariationalDensity_{t-1 \mid t}^{0:T} &=
\argmax[\VariationalDensity_t \in \VariationalFamily_t,\, \VariationalDensity_{t-1\mid t} \in \VariationalFamily_{t-1\mid t}]
\int \log \Big\{\frac{\UnnormalizedDensity_{t \mid t-1}^t \ForwardRepresenter_{t-1}^{0:t-1}}
{\VariationalDensity_{t-1 \mid t} \VariationalDensity_t}\Big\}
 \VariationalDensity_{t-1 \mid t} \VariationalDensity_t \dif \statevar_{t-1:t} \\
\log \ForwardRepresenter_t^{0:t} &=
\int \log \Big\{\frac{ \UnnormalizedDensity_{t \mid t-1}^t \ForwardRepresenter_{t-1}^{0:t-1}}{\VariationalDensity_{t-1 \mid t}^{0:T} }\Big\}
\VariationalDensity_{t-1 \mid t}^{0:T} \dif \statevar_{t-1}.
\end{align}
\end{subequations}
\end{definition}
The procedure of \cref{def:sub-optimal-variational-filter-1} is indeed sub-optimal but new definition of $\ForwardRepresenter_t^{0:t}$ in fact also provides a lower bound on the unnormalized filtering density $\UnnormalizedDensity_t^{0:t}$
-- the argument is exactly the same as for \cref{cor:filter-lower-bound}.
It is just not necessarily the optimal bound obtainable with respect to the chosen variational family.
Furthermore, the sub-optimal variational filter serves as a reasonable method of initializing the fixed-point iteration of \cref{prop:iterative-forward-smoother}.

\section{On the intractability of the value functionals}\label{sec:intractable-value-functionals}
The fact that the value functional recursions of \cref{thm:bayesian-backward-recursion}, \cref{thm:bayesian-forward-recursion},
and \cref{def:sub-optimal-variational-filter-1} are in general intractable has hitherto not been discussed.
However, $\ForwardRepresenter_t$ ought to be an unnormalized density (see \cref{cor:filter-lower-bound}).
Therefore, some care needs to taken to ensure that $\ForwardRepresenter_t$ remains normalizable throughout recursion.
Here this issue shall not be examined in exhaustive detail, there is simply too many possible approaches that will have to be explored elsewhere.
However, there is a simple approach approach will be offered.
Namely, by \cref{def:sub-optimal-variational-filter-1}, the normalization constant of $\ForwardRepresenter_t$ has a lower bound according to
\begin{equation}
\begin{split}
\log\int \ForwardRepresenter_t(\statevar_t) \dif \statevar_t
&\geq   \int \log \Big\{\frac{\ForwardRepresenter_t(\statevar_t)}{\VariationalDensity_t(\statevar_t)}\Big\} \VariationalDensity_t^{0:t}(\statevar_t) \dif \statevar_t \\
&= \int \log \Big\{\frac{ \UnnormalizedDensity_{t \mid t-1}^t(\statevar_t \mid \statevar_{t-1}) \ForwardRepresenter_{t-1}(\statevar_{t-1})}
{\VariationalDensity_{t-1 \mid t}^{0:T}(\statevar_{t-1} \mid \statevar_t) \VariationalDensity_t^{0:t}(\statevar_t)}\Big\}
\VariationalDensity_{t-1 \mid t}^{0:T}(\statevar_{t-1} \mid \statevar_t)  \VariationalDensity_t^{0:t}(\statevar_t) \dif \statevar_{t-1:t}.
\end{split}
\end{equation}
This then serves as an approximation of the log-normalizing constant of $\ForwardRepresenter_t$,
combining this with the fact that $\VariationalDensity_t^{0:t}$ is the best approximation of the normalization of $\ForwardRepresenter_t$ gives the following approximation.
\begin{approximation}[Variational approximation of the forward value functional I]
\label{app:forward-value-functional-approximation-1}
The forward value functional is approximately given by
\begin{subequations}
\begin{align}
\log \hat{\kappa}_t &=  \int \log \Big\{\frac{ \UnnormalizedDensity_{t \mid t-1}^t \ForwardRepresenter_{t-1}}{\VariationalDensity_{t-1 \mid t}^{0:T} \VariationalDensity_t^{0:t}(\statevar_t)}\Big\}
\VariationalDensity_{t-1 \mid t}^{0:T} \VariationalDensity_t^{0:t}(\statevar_t) \dif \statevar_{t-1:t} \\
\log \ForwardRepresenter_t &\approx \log \hat{\kappa}_t + \log \VariationalDensity_t^{0:t}(\statevar_t).
\end{align}
\end{subequations}
\end{approximation}

The filter proposed by \cite{Courts2021} is obtained when the sub-optimal variational filter of \cref{def:sub-optimal-variational-filter-1} is combined with \cref{app:forward-value-functional-approximation-1}.
Namely, by plugging in \cref{app:forward-value-functional-approximation-1} into \cref{def:sub-optimal-variational-filter-1}, the maximization problem of the filter update reads
\begin{equation}
\underset{\VariationalDensity_{t-1} \in \VariationalFamily_{t-1},\, \VariationalDensity_{t \mid t-1} \in \VariationalFamily_{t \mid t-1}}{\max}\,
\int \log \Big\{\frac{\UnnormalizedDensity_{t \mid t-1}^t \VariationalDensity_{t-1}^{0:t-1}}
{\VariationalDensity_{t \mid t-1} \VariationalDensity_{t-1}}\Big\}
 \VariationalDensity_{t-1 \mid t} \VariationalDensity_t \dif \statevar_{t-1:t},
\end{equation}
where Bayes' rule was used, $\VariationalDensity_{t-1 \mid t} \VariationalDensity_t = \VariationalDensity_{t \mid t-1} \VariationalDensity_{t-1}$.
This is precisely algorithm 2 of \cite{Courts2021}.

\section{Related work}\label{sec:related-work}

Recursive formulations have previously been studied by primarily \citet{Weber2015} and \citet{Campbell2021} from the reinforcement learning perspective.
These recursions are developed in terms of certain \emph{value functions} that of course are related to the present notion of \emph{value functionals}.
More precisely, \citet{Weber2015} defines a backward value function by
\begin{equation*}
\tilde{\BackwardValueFunction}_t(\statevar_t) = \int \log \Big\{ \frac{\UnnormalizedDensity_{t+1:T\mid t}^{t+1:T}(\statevar_{t+1:T} \mid \statevar_t)}{\VariationalDensity_{t+1:T\mid t}(\statevar_{t+1:T} \mid \statevar_t)} \Big\}
\VariationalDensity_{t+1:T\mid t}(\statevar_{t+1:T} \mid \statevar_t) \dif \statevar_{t+1:T}.
\end{equation*}
The following relation is then formally identified (see \cref{def:variational-backward-value})
\begin{equation*}
\tilde{\BackwardValueFunction}_t(\statevar_t) = \log \BackwardRepresenter_t(\statevar_t).
\end{equation*}
However, the relation between $\BackwardRepresenter_t$ and $\ObsDensity_{t+1:T\mid t}$ provided by \cref{cor:likelihood-lower-bound} was omitted by \citet{Weber2015} and so was the fixed-point equations of \cref{prop:iterative-forward-smoother}.
On the other hand, \citet{Campbell2021} defines a forward value function by
\begin{equation*}
\tilde{\ForwardValueFunction}_t(\statevar_t) = \int \log \Big\{ \frac{\UnnormalizedDensity_{0:t}^{0:t}(\statevar_{0:t})}{\VariationalDensity_{0:t}(\statevar_{0:t})} \Big\}
\VariationalDensity_{0:t-1\mid t}(\statevar_{0:t-1} \mid \statevar_t) \dif \statevar_{0:t-1}.
\end{equation*}
In view of \cref{def:variational-forward-value} and \cref{thm:variational-forward-hjb} this value function may formally be identified as
\begin{equation*}
\tilde{\ForwardValueFunction}_t(\statevar_t) = \log \Big\{ \frac{\ForwardRepresenter_t(\statevar_t)}{\VariationalDensity_t(\statevar_t)}\Big\}.
\end{equation*}
However, the relation between $\UnnormalizedDensity_t^{0:t}$ and $\ForwardRepresenter_t$ established by \cref{cor:filter-lower-bound} was not provided by \citet{Campbell2021},
nor was the fixed-point equations \cref{prop:iterative-forward-smoother}.
Furthermore, they put forth the following value function recursion
\begin{equation}\label{eq:campbell-recursion}
\begin{split}
\tilde{\ForwardValueFunction}_{t+1}(\statevar_{t+1}) &= \int \log \Big\{ \frac{\UnnormalizedDensity_{t+1\mid t}^{t+1}(\statevar_{t+1} \mid \statevar_t) \VariationalDensity_t(\statevar_t)}
{\VariationalDensity_{t\mid t+1}(\statevar_t \mid \statevar_{t+1}) \VariationalDensity_{t+1}(\statevar_{t+1})} \Big\} \VariationalDensity_{t\mid t+1}(\statevar_t \mid \statevar_{t+1}) \dif \statevar_t \\
&\quad + \int \tilde{\ForwardValueFunction}_t(\statevar_t) \VariationalDensity_{t\mid t+1}(\statevar_t \mid \statevar_{t+1}) \dif \statevar_t.
\end{split}
\end{equation}
Now if $\VariationalDensity_t$ actually satisfies the following constraint
\begin{equation*}
\VariationalDensity_t(\statevar_t) = \int \VariationalDensity_{t\mid t+1}(\statevar_t \mid \statevar_{t+1}) \VariationalDensity_{t+1}(\statevar_{t+1}),
\end{equation*}
then the value function recursion of \eqref{eq:campbell-recursion} is formally equivalent to \cref{thm:vjgm-forward-value-functional}, namely
\begin{equation*}
\tilde{\ForwardValueFunction}_{t+1}(\statevar_{t+1}) = \int \log \Big\{ \frac{\UnnormalizedDensity_{t+1\mid t}^{t+1}(\statevar_{t+1} \mid \statevar_t) \ForwardRepresenter_t(\statevar_t)}
{\VariationalDensity_{t\mid t+1}(\statevar_t \mid \statevar_{t+1}) \VariationalDensity_{t+1}(\statevar_{t+1})} \Big\} \VariationalDensity_{t\mid t+1}(\statevar_t \mid \statevar_{t+1}) \dif \statevar_t.
\end{equation*}
However, this is not the approach suggested by \citet{Campbell2021}.
Instead they identify $\VariationalDensity_t$ with their corresponding filtering distribution $\VariationalDensity_t^{0:t}$.
Notably, the present approach circumvents this issue by recursing over $\ForwardRepresenter_t$ rather than $\tilde{\ForwardValueFunction}_t$.
This also means that the present notion of variational filtering is different from that of \citet{Campbell2021}.

A few days later after the first pre-print of the present manuscript was released, another approach appeared by \citep{Abdulsamad2025}.
Where they study iterative recursive variational Gaussian smoothing under entropic regularization using a Lagrange multiplier approach rather than dynamic programming.
Also, rather than imposing a constraint class $\VariationalFamily_{0:T}$ explicitly they ``close'' the value functional recursions by a variety of quadratic approximations.

\section{Case study: Jump Gauss--Markov systems}\label{sec:jump-gauss-markov}
In this section the variational forward method is studied in jump Gauss--Markov systems
under a certain factored approximation, which leads to an alternate coordinate ascent method for computing  the transition kernels.
More specifically, consider the following Markov chain
\begin{subequations}
\begin{align}
\auxvar_0 &\sim \lambda_0(\auxvar_0) \\
\auxvar_t \mid \auxvar_{t-1} &\sim \lambda_{t \mid t-1}(\auxvar_t \mid \auxvar_{t-1}),
\end{align}
\end{subequations}
where $\auxvar_t$ takes values in $\{1,\ldots, M\}$.
The process $\auxvar_t$ drives two other processes $\statevar_t$ and $\obsvar_t$ according to
\begin{subequations}
\begin{align}
\statevar_0 \mid \auxvar_0 &\sim \rho_0(\statevar_0 \mid \auxvar_0) \\
\statevar_t \mid \statevar_{t-1}\,,\auxvar_{t-1} &\sim \rho_{t\mid t-1}(\statevar_t \mid \statevar_{t-1}, \auxvar_{t-1}) \\
\obsvar_t \mid \statevar_t\,, \auxvar_t &\sim h_{t\mid t}(\obsvar_t \mid \statevar_t, \auxvar_t),
\end{align}
\end{subequations}
Additionally, a conditionally Gaussian assumption is imposed on $\statevar_t$ and $\obsvar_t$, namely
\begin{subequations}
\begin{align}
\rho_0(\statevar_0\mid \auxvar_0) &= \gaussian(\statevar_0; \StateMean_0(\auxvar_0), \StateCov_0(\auxvar_0) ) \\
\StateCondMean_{t\mid t-1}(\statevar_{t-1}, \auxvar_{t-1}) &= \TransitionMatrix_{t, t-1}(\auxvar_{t-1})\statevar_{t-1} + \StateControl_{t,t-1}(\auxvar_{t-1})\\
\rho_{t\mid t-1}(\statevar_t \mid \statevar_{t-1}, \auxvar_{t-1}) &=
\gaussian(\statevar_t; \StateCondMean_{t\mid t-1}(\statevar_{t-1}, \auxvar_{t-1}), \StateCondCov_{t\mid t-1}(\auxvar_{t-1}) ) \\
\ObsCondMean_{t\mid t}(\statevar_t, \auxvar_t) &= \ObsMatrix_{t\mid t}(\auxvar_t) \statevar_t + \ObsBias_{t\mid t}(\auxvar_t) \\
\ObsDensity_{t\mid t}(\obsvar_t \mid \statevar_t, \auxvar_t) &= \gaussian(\obsvar_t; \ObsCondMean_{t\mid t}(\statevar_t, \auxvar_t), \ObsCondCov_{t\mid t}(\auxvar_t)).
\end{align}
\end{subequations}
Taken together, the triplet $(\auxvar_t, \statevar_t, \obsvar_t)$ forms a jump Gauss--Markov system and the goal here is to infer the complete hidden state $(\auxvar_t, \statevar_t)$
given the observed sequence $\obsvar_{0:T}$.
The transition kernel of the complete hidden state is then
\begin{equation}
\StateDensity_{t \mid t-1}(\statevar_t, \auxvar_t \mid \statevar_{t-1}, \auxvar_{t-1} ) = \rho_{t\mid t-1}(\statevar_t \mid \statevar_{t-1}, \auxvar_{t-1}) \lambda_{t \mid t-1}(\auxvar_t \mid \auxvar_{t-1}).
\end{equation}
In this section the variational state estimation methodology is applied under a factorized variational approximation according to
\begin{equation}\label{eq:jump-gauss-markov-free-form-factorization}
\VariationalDensity_{0:T}(\statevar_{0:T}, \auxvar_{0:T}) =  f_{0:T}(\auxvar_{0:T}) g_{0:T}(\statevar_{0:T}).
\end{equation}
The aim of this section is to show, under the factorized class \eqref{eq:jump-gauss-markov-free-form-factorization}, that the forward value functional satisfy the following theorem.

\begin{theorem}[Tractable forward value functional]
\label{thm:vjgm-forward-value-functional}
The forward value functional is for $0 \leq t \leq T$ respectively given by
\begin{equation}\label{vjgm:forward_value}
\log \ForwardRepresenter_t(\statevar_t, \auxvar_t) = \log \kappa_t +  \log f^\star_t(\auxvar_t) + \log g^\star_t(\statevar_t \mid \auxvar_t),
\end{equation}
where $g^\star_t$ is given by
\begin{equation}
g^\star_t(\statevar_t \mid \auxvar_t) = \gaussian(\statevar_t; \StateMean^\star_t(\auxvar_t), \StateCov^\star_t(\auxvar_t) ).
\end{equation}
Additionally, the optimal $g_{0:T}$ is Gauss--Markov.
\end{theorem}
Demonstrating the veracity of \cref{thm:vjgm-forward-value-functional} and developing the value functional recurrences will be done in several steps.
Before doing so, the filtering update will be established.
Namely, let $\xi_t$ be the negative relative entropy between $g_t$ and $g_t^\star$, that is
\begin{equation}
\begin{split}
\log \xi_t(\auxvar_t) &= \int \log \Big\{\frac{g^\star_t(\statevar_t \mid \auxvar_t)}
{g_t(\statevar_t)} \Big\} g_t(\statevar_t)\dif \statevar_t.
\end{split}
\end{equation}
Then in view of \cref{def:sub-optimal-variational-filter-1}, the filtering distributions $g_t$ and $f_t$ maximize:
\begin{equation*}
\sum_{\auxvar_t} \Big( \int \log \Big\{ \frac{\ForwardRepresenter_t(\statevar_t, \auxvar_t)}{g_t(\statevar_t) f_t(\auxvar_t)} \Big\}  g_t(\statevar_t) \dif \statevar_t \Big) f_t(\auxvar_t)
= \sum_{\auxvar_t} \log \Big\{ \frac{\xi_t(\auxvar_t)f_t^\star(\auxvar_t)}{f_t(\auxvar_t)} \Big\} f_t(\auxvar_t) + \log \kappa_t.
\end{equation*}
Clearly, the stationary point with respect to $f_t$ is given by $f_t \propto \xi_t f_t^\star$.
Similarly, the stationary point with respect to $g_t$ satisfy
\begin{equation*}
g_t(\statevar_t) \propto \exp \Big( \sum_{\auxvar_t} \log\{\ForwardRepresenter_t(\statevar_t, \auxvar_t)\}  f_t(\auxvar_t) \Big)
\end{equation*}
and by \cref{thm:vjgm-forward-value-functional} $g_t$ is then log-quadratic in $\statevar_t$.
That is to say Gaussian, the parameters of which are obtained by simple algebra.
The following theorem is obtained.

\begin{theorem}
\label{thm:vjgm-update}
If \cref{thm:vjgm-forward-value-functional} hold, then the filtering distributions $g_t$ and $f_t$ satisfy the following fixed-point equations
\begin{subequations}
\begin{align}
f_t(\auxvar_t) &= \frac{\xi_t(\auxvar_t)f_t^\star(\auxvar_t) }{ \sum_{\auxvar} \xi_t(\auxvar) f_t^\star(\auxvar)} \\
\StateCov_t^{-1} &= \sum_{\auxvar_t} \{\StateCov^\star_t(\auxvar_t)\}^{-1}  f_t(\auxvar_t) \\
\StateCov_t^{-1}\StateMean_t   &=  \sum_{\auxvar_t} \{\StateCov^\star_t(\auxvar_t)\}^{-1} \StateMean^\star_t(\auxvar_t) f_t(\auxvar_t).
\end{align}
\end{subequations}
Additionally, the marginal likelihood is lower bounded by
\begin{equation}
\log h_{0:t} \geq  \log\Big\{ \sum_{\auxvar} \xi_t(\auxvar)f_t^\star(\auxvar)\Big\} + \log \kappa_t.
\end{equation}
\end{theorem}

\subsection{The variational forward method}

For the induction step, assume that \eqref{vjgm:forward_value} holds at $t-1$ in the following.
Start by applying Bayes' rule to $\lambda_{t \mid t-1}$ and $f^\star_{t-1}$,
which gives the following identity
\begin{equation}\label{eq:jump-gauss-markov-model-prediction}
\lambda_{t\mid t-1}(\auxvar_t \mid \auxvar_{t-1}) f^\star_{t-1}(\auxvar_{t-1}) =  f_t^{\mathsf{p}}(\auxvar_t) f_{t-1\mid t}^\dagger(\auxvar_{t-1} \mid \auxvar_t)
\end{equation}
Similarly, the following identity also hold by Bayes' rule
\begin{subequations}\label{eq:jump-gauss-markov-state-prediction}
\begin{align}
\rho_{t\mid t-1}(\statevar_t \mid \statevar_{t-1}, \auxvar_{t-1}) g^\star_{t-1}(\statevar_{t-1} \mid \auxvar_{t-1}) &=
\tilde{g}^{\mathsf{p}}_t(\statevar_t\mid \auxvar_{t-1}) \tilde{g}_{t-1 \mid t}(\statevar_{t-1} \mid \statevar_t, \auxvar_{t-1}) \\
\tilde{g}^{\mathsf{p}}_t(\statevar_t\mid \auxvar_{t-1}) &= \gaussian(\statevar_t; \tilde{\StateMean}^{\mathsf{p}}_t(\auxvar_{t-1}), \tilde{\StateCov}^{\mathsf{p}}_t(\auxvar_{t-1})) \\
\tilde{g}_{t-1 \mid t}(\statevar_{t-1} \mid \statevar_t, \auxvar_{t-1}) &=
\gaussian(\statevar_{t-1}; \tilde{\StateCondMean}_{t-1\mid t}(\statevar_t, \auxvar_{t-1}), \tilde{\StateCondCov}_{t-1\mid t}(\auxvar_{t-1})),
\end{align}
\end{subequations}
where $\tilde{g}^{\mathsf{p}}_t$ , and $\tilde{g}_{t-1 \mid t}$ are obtained by Gaussian marginalization and conditioning, respectively.
Inserting this into \cref{thm:variational-forward-hjb} gives the following result.
\begin{lemma}\label{lem:forward-value-functional-1}
If \eqref{vjgm:forward_value} hold at $t-1$,
then the forward value functional at time $t$ is given by
\begin{equation*}
\begin{split}
&\log \ForwardRepresenter_t(\statevar_t, \auxvar_t) = \log f_t^{\mathsf{p}}(\auxvar_t) +
\sum_{\auxvar_{t-1}} \log \Big\{ \frac{f_{t-1\mid t}^\dagger(\auxvar_{t-1} \mid \auxvar_t)}{f_{t-1\mid t}(\auxvar_{t-1} \mid \auxvar_t)}\Big\} f_{t-1\mid t}(\auxvar_{t-1} \mid \auxvar_t) + \log \ObsDensity_{t\mid t}(\statevar_t, \auxvar_t)\\
& + \sum_{\auxvar_{t-1}}
\Big( \int \log \Big\{\frac{\tilde{g}_{t-1 \mid t}(\statevar_{t-1} \mid \statevar_t, \auxvar_{t-1})}{g_{t-1\mid t}(\statevar_{t-1}\mid \statevar_t)}\Big\} g_{t-1\mid t}(\statevar_{t-1}\mid \statevar_t) \dif \statevar_{t-1}\Big)
f_{t-1\mid t}(\auxvar_{t-1} \mid \auxvar_t) \\
&+ \log \kappa_{t-1} + \sum_{\auxvar_{t-1}} \log\{\tilde{g}^{\mathsf{p}}_t(\statevar_t\mid \auxvar_{t-1})\} f_{t-1\mid t}(\auxvar_{t-1} \mid \auxvar_t).
\end{split}
\end{equation*}
\end{lemma}
\Cref{lem:forward-value-functional-1} will be useful for obtaining a fixed-point equation for the reverse-time transition kernels in the sequel.
Before doing so, define the likelihood $\eta_t$,  the Gaussian transiton kernel $g^\dagger_{t-1\mid t}$, and the Gaussian density $g^{\mathsf{p}}_t$ by
\footnote{A convenient way of organizing of $\eta_t$, $g^\dagger_{t-1\mid t}$, and $g^{\mathsf{p}}_t$ computation is provided by \cref{lem:average-log-gaussian} in \cref{app:vjgm}}
\begin{equation}
\begin{split}
&\log\{\eta_t(\auxvar_t) g^\dagger_{t-1\mid t}(\statevar_{t-1} \mid \statevar_t, \auxvar_t) g^{\mathsf{p}}_t(\statevar_t \mid \auxvar_t)\} \\
&\quad= \sum_{\auxvar_{t-1}} \log\{\tilde{g}^{\mathsf{p}}_t(\statevar_t\mid \auxvar_{t-1}) \tilde{g}_{t-1 \mid t}(\statevar_{t-1} \mid \statevar_t, \auxvar_{t-1}) \} f_{t-1\mid t}(\auxvar_{t-1} \mid \auxvar_t).
\end{split}
\end{equation}
Additionally, define the likelihoods $\zeta_t^\dagger$, $\eta_t^\dagger$ and $\ObsDensity_{t\mid t}^\dagger$ by
\footnote{
A convenient way of organizing the computation of $\eta_t^\dagger$ and $\ObsDensity_{t\mid t}^\dagger$ is provided by \cref{lem:relative-entropy-likelihood} in \cref{app:vjgm}.
}
\begin{subequations}
\begin{align}
\log \zeta_t^\dagger(\auxvar_t) &= \sum_{\auxvar_{t-1}} \log \Big\{ \frac{f_{t-1\mid t}^\dagger(\auxvar_{t-1} \mid \auxvar_t)}{f_{t-1\mid t}(\auxvar_{t-1} \mid \auxvar_t)}\Big\}
 f_{t-1\mid t}(\auxvar_{t-1} \mid \auxvar_t) \\
\log\{\eta_t^\dagger(\auxvar_t) \ObsDensity_{t\mid t}^\dagger(\statevar_t, \auxvar_t)\} &=
\int \log \Big\{\frac{g^\dagger_{t-1\mid t}(\statevar_{t-1} \mid \statevar_t, \auxvar_t)}{g_{t-1\mid t}(\statevar_{t-1}\mid \statevar_t)}\Big\}
g_{t-1\mid t}(\statevar_{t-1}\mid \statevar_t) \dif \statevar_{t-1},
\end{align}
\end{subequations}
where it can be seen that $\ObsDensity_{t\mid t}^\dagger$ is indeed a Gaussian likelihood.
Then by successive substitutions, the forward value functional is given by
\begin{equation*}
\begin{split}
\log \ForwardRepresenter_t(\statevar_t, \auxvar_t) &= \log \{ \eta_t^\dagger(\auxvar_t)  \zeta_t^\dagger(\auxvar_t) \eta_t(\auxvar_t)  f_t^{\mathsf{p}}(\auxvar_t)\} \\
&\quad + \log \{\ObsDensity_{t\mid t}^\dagger(\statevar_t, \auxvar_t) \ObsDensity_{t\mid t}(\statevar_t, \auxvar_t)  g^{\mathsf{p}}_t(\statevar_t \mid \auxvar_t)\}
+ \log \kappa_{t-1}.
\end{split}
\end{equation*}
The forward value functional update is now readily obtained and the proof of \cref{thm:vjgm-forward-value-functional} is complete.
\begin{proposition}[Forward value functional update]
Given that $\ForwardRepresenter_{t-1}$ is of the form stated in \cref{thm:vjgm-forward-value-functional}, then so is $\ForwardRepresenter_t$.
The update is given by the following relations:
\begin{subequations}
\begin{align}
\bar{\ObsDensity}_{t\mid t}(\auxvar_t) g^\star_t(\statevar_t\mid \auxvar_t) &=
\ObsDensity_{t\mid t}^\dagger(\statevar_t, \auxvar_t) \ObsDensity_{t\mid t}(\statevar_t, \auxvar_t)  g^{\mathsf{p}}_t(\statevar_t \mid \auxvar_t)\\
\kappa_{t\mid t-1} &= \sum_{\auxvar_t} \bar{\ObsDensity}_{t\mid t}(\auxvar_t)\eta_t^\dagger(\auxvar_t)  \zeta_t^\dagger(\auxvar_t) \tilde{\eta}_t(\auxvar_t)  f_t^{\mathsf{p}}(\auxvar_t) \\
f_t^\star(\auxvar_t) &= \kappa_{t\mid t-1}^{-1} \bar{\ObsDensity}_{t\mid t}(\auxvar_t)\eta_t^\dagger(\auxvar_t)  \zeta_t^\dagger(\auxvar_t) \tilde{\eta}_t(\auxvar_t)  f_t^{\mathsf{p}}(\auxvar_t)\\
\log \kappa_t &= \log \kappa_{t-1} + \log \kappa_{t\mid t-1},
\end{align}
\end{subequations}
where the first relation is Bayes' rule.
\end{proposition}

\subsubsection{Fixed-point equations for the transition kernels}
It remains to obtain the fixed-point equations for the reverse-time transition kernels $g_{t-1\mid t}$ and $f_{t-1\mid t}$.
They are given by as the maximizers of
\begin{equation*}
\sum_{\auxvar_t} \Big( \int \log \ForwardRepresenter_t(\statevar_t, \auxvar_t) g_t(\statevar_t) \dif \statevar_t \Big) f_t(\auxvar_t).
\end{equation*}
In view of \cref{lem:forward-value-functional-1}, define $\zeta_{t-1}$ by
\begin{equation}
\begin{split}
\log \zeta_{t-1}(\auxvar_{t-1}) &= \int
\Big( \int \log \Big\{\frac{\tilde{g}_{t-1 \mid t}(\statevar_{t-1} \mid \statevar_t, \auxvar_{t-1})}{g_{t-1\mid t}(\statevar_{t-1}\mid \statevar_t)}\Big\}
g_{t-1\mid t}(\statevar_{t-1}\mid \statevar_t) \dif \statevar_{t-1}\Big) g_t(\statevar_t) \dif \statevar_t \\
&\quad + \int  \log\{\tilde{g}^{\mathsf{p}}_t(\statevar_t\mid \auxvar_{t-1})\} g_t(\statevar_t) \dif \statevar_t,
\end{split}
\end{equation}
then it is immediately clear that $f_{t-1\mid t} \propto \zeta_{t-1} f_{t-1\mid t}^\dagger$.
Additionally, by \cref{lem:forward-value-functional-1} it is clear that $g_{t-1 \mid t}$ is given as the maximizer of
\begin{equation*}
\sum_{\auxvar_t, \auxvar_{t-1}}
\Big( \int \log \Big\{\frac{\tilde{g}_{t-1 \mid t}(\statevar_{t-1} \mid \statevar_t, \auxvar_{t-1})}{g_{t-1\mid t}(\statevar_{t-1}\mid \statevar_t)}\Big\} g_{t-1\mid t}(\statevar_{t-1}\mid \statevar_t) \dif \statevar_{t-1}\Big)
f_{t-1\mid t}(\auxvar_{t-1} \mid \auxvar_t) f_t(\auxvar_t).
\end{equation*}
Consequently, the reverse-time transition kernel $g_{t-1\mid t}$  is clearly given by
\begin{equation}
g_{t-1\mid t}(\statevar_{t-1}\mid \statevar_t) \propto
\exp \Big(  \sum_{\auxvar_t, \auxvar_{t-1}} \log\{ \tilde{g}_{t-1 \mid t}(\statevar_{t-1} \mid \statevar_t, \auxvar_{t-1}) \} f_{t-1\mid t}(\auxvar_{t-1} \mid \auxvar_t) f_t(\auxvar_t) \Big)
\end{equation}
and since $\tilde{g}_{t-1 \mid t}$ is log-quadratic in $\statevar_{t-1}$ and $\statevar_t$ then  so is $g_{t-1\mid t}$, Gaussian that is.

\begin{proposition}[Fixed-point equations fo reverse-time transition kernels]
The reverse-time transiton kernels satisfy the following fixed-point equations
\begin{align}
f_{t-1\mid t}(\auxvar_{t-1} \mid \auxvar_t) &=
\frac{\zeta_{t-1}(\auxvar_{t-1}) f_{t-1\mid t}^\dagger(\auxvar_{t-1} \mid \auxvar_t) }{ \sum_{\auxvar_{t-1}} \zeta_{t-1}(\auxvar_{t-1}) f_{t-1\mid t}^\dagger(\auxvar_{t-1} \mid \auxvar_t)} \\
Q_{t-1\mid t}^{-1} &= \sum_{\auxvar_{t-1},\auxvar_t} \tilde{Q}_{t-1\mid t}^{-1}(\auxvar_{t-1}) f_{t-1\mid t}(\auxvar_{t-1} \mid \auxvar_t) f_t(\auxvar_t) \\
Q_{t-1\mid t}^{-1} a_{t-1\mid t}(\statevar) &=    \sum_{\auxvar_{t-1},\auxvar_t} \tilde{Q}_{t-1\mid t}^{-1}(\auxvar_{t-1}) \StateCondMean_{t-1\mid t}(\statevar, \auxvar_{t-1})
f_{t-1\mid t}(\auxvar_{t-1} \mid \auxvar_t) f_t(\auxvar_t) \\
g_{t-1\mid t}(\statevar_{t-1} \mid \statevar_t) &= \gaussian(a_{t-1\mid t}(\statevar), Q_{t-1\mid t}).
\end{align}
\end{proposition}

\subsection{Simulation study: an autoregressive staircase}

\begin{figure}[t!]
\centering
\includegraphics{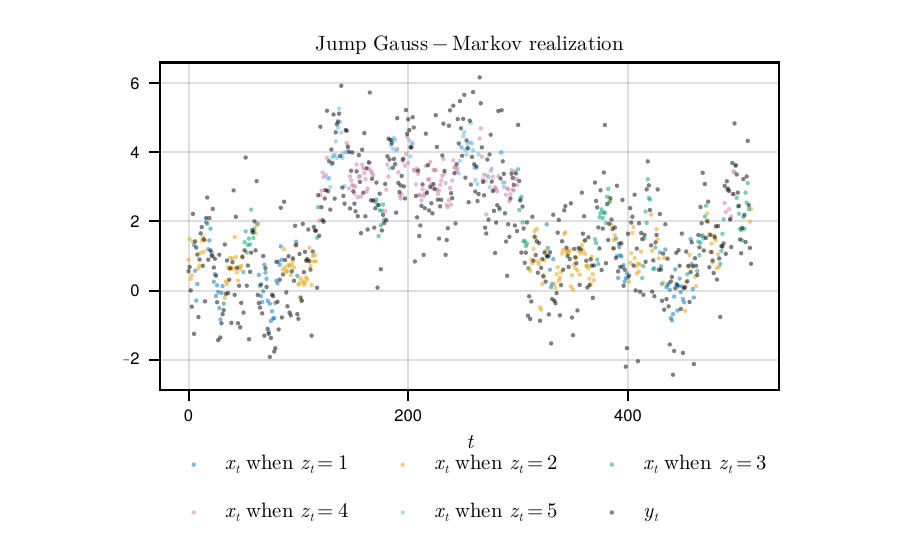}
\caption{
One realization of the jump Gauss--Markov system.
The markers for $\statevar_t$ are colored according to the corresponding value of $\auxvar_t$.
}\label{fig:staircase_realization}
\end{figure}

The model index, $\auxvar_t$, starts at the uniform distribution on $\{1, \ldots, M\}$ and then stays in its current state with probability $p$ or jumps to any adjacent state with equal probability.
The Markov chain is then specified by the following initial and transition probabilities,
\begin{subequations}
\begin{align}
\lambda_0(\auxvar) &= \frac{1}{M} \\
\lambda_{t\mid t-1}(\auxvar \mid 1) &= p \delta_{\auxvar, 1} + (1 - p) \delta_{\auxvar, 2} \\
\lambda_{t\mid t-1}(\auxvar \mid \auxvar') &= p \delta_{\auxvar, \auxvar'} + \frac{1 - p}{2} ( \delta_{\auxvar-1, \auxvar'} + \delta_{\auxvar+1, \auxvar'} ), \quad \auxvar' = 2, \ldots, M-1,\\
\lambda_{t\mid t-1}(\auxvar \mid M) &= p \delta_{\auxvar, M} + (1 - p) \delta_{\auxvar, M-1} \\
\end{align}
\end{subequations}
where $\delta$ denotes Kronecker's delta.
The state variable switches between $M$ different first order, stationary autoregressive processes with parameters given by
\begin{subequations}
\begin{align}
\StateMean_0(\auxvar) &= z-1 \\
\StateCondCov_{t\mid t-1} &= (1 - \abs{\phi_0}^2) \Sigma_0(\auxvar) \\
\StateCondMean_{t\mid t-1}(\statevar, \auxvar) &= \phi_0 x + (1 - \phi_0) \mu_0(\auxvar),
\end{align}
\end{subequations}
where $\StateMean_0(\auxvar)$ and $\Sigma_0(\auxvar)$ are the stationary means and variances, repectively.
Lastly, the parameters of the observation model are given by
\begin{subequations}
\begin{align}
\ObsCondMean_{t\mid t}(\statevar, \auxvar) &= x \\
\ObsCondCov_{t\mid t}(\statevar, \auxvar)  &= R.
\end{align}
\end{subequations}

There are various recursive approaches to computing filtering densities and smoothing marginals in jump Gauss--Markov systems \citep{BarShalom2004,Helmick1995,Balenzuela2022}.
However, to the authors knowledge, none that can recursively compute a posterior over the entire path $(\statevar_{0:T}, \auxvar_{0:T})$.
In any case, the posterior obtained from \cref{prop:iterative-forward-smoother} is optimal in relative entropy sense with respect to the chosen variational family.
Thus instead the sub-optimal variational filter (VJGM(0)) is compared to an interacting multiple model filter with $M$ components (IMM) \citep{BarShalom2004},
to ensure the notion of sub-optimal vairational filtering is reasonable.
Additionally, (VJGM(0)) is used to initialize the fixed-point iteration of \cref{prop:iterative-forward-smoother}, which runs for 10 iterations (VJGM(10)).
\footnote{
It seems to always converge in less than 10 iterations in this example.
}
The posteriors obtained from (VJGM(0)) and (VJGM(10)) are then compared.
The parameters are set to $\StateMean_0(\auxvar) = \auxvar - 1$, $\StateCov_0(\auxvar) = 0.5^2$,
$\phi_0 = 0.5$, $M = 4$, $p = 0.9$, $R = 1$, and the number of time points is set to $T = 2^9+1$.
A realization of the system is shown in \cref{fig:staircase_realization}.

\begin{figure}[t!]
\centering
\includegraphics{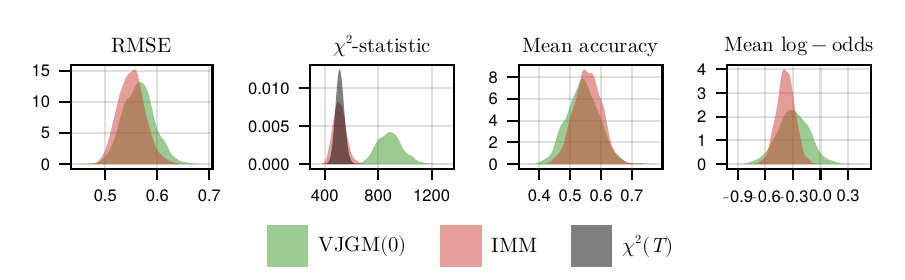}
\caption{
Performance comparison of the filtering distributions produced by VJGM(0) and IMM.
For the $\chi^2$-statistic, the corresponding ground-truth density is shown.
It can be seen that their performance is quite similar across metrics except for the $\chi^2$-square statistic,
where IMM significantly outperforms VJGM(0).
}
\label{fig:vjgm-vs-imm}
\end{figure}

The comparisons are carried out over 1000 Monte Carlo trials where
the root mean-square errors in estimating $\statevar_t$ are recorded along with
the mean accuracy and mean log-odds of the correct $\auxvar_t$ under the filtering distribution when comparing VJGM(0) with IMM and under the smoothing distribution when comparing VJGM(0) with VJGM(10).
Additionally, in order to assess the calibration quality in the estimates of $\statevar_t$ the sum of $\chi^2$-statistics is recorded when comparing VJGM(0) with IMM and the $\chi^2$-statistic of the entire
sequence $\statevar_{0:T}$ when comparing VJGM(0) and VJGM(10).
\footnote{
When the Gaussian assumption hold both $\chi^2$ metrics are realizations from $\chi^2(T+1)$.
}

The results of the comparison between VJGM(0) and IMM are shown in \cref{fig:vjgm-vs-imm}.
It can be seen that their performance is similar across all metrics except the $\chi^2$-statistic,
where the uncertainty in $\statevar_t$ produced by VJGM(0) is very overconfident.
This might be due to the well documented mode seeking behavior of variational inference \citep{Bishop2006}.

\begin{figure}
\centering
\includegraphics{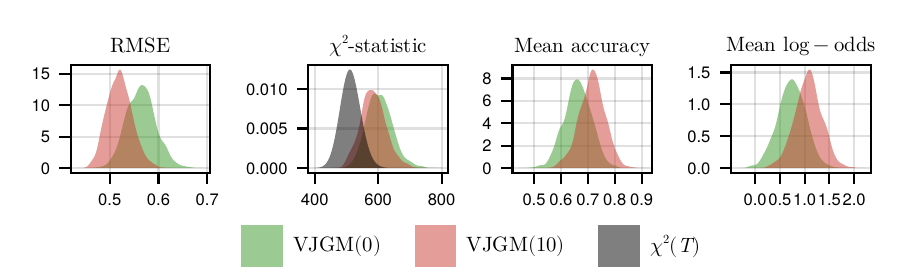}
\caption{
Performance comparison of posterior distributions over the path $(\statevar_{0:T}, \auxvar_{0:T})$ produced by VJGM(0) and VJGM(10).
For the $\chi^2$-statistic, the corresponding ground-truth density is shown.
It can be seen that VJGM(10) offers a moderate improvement over VJGM(0) in terms of estimation accuracy, calibration, and classification accuracy.
}\label{fig:vjgm-vs-vjgm}
\end{figure}

The result of the comparison between VJGM(0) and VJGM(10) are shown in \cref{fig:vjgm-vs-vjgm}.
Here it can be seen that while VJGM(0) produces a posterior of reasonable quality,
actually carrying out the fixed-point iteration of \cref{prop:iterative-forward-smoother} can bring signifcant improvements.
Comparing to the filtering results of \cref{fig:vjgm-vs-imm}, it can also be seem that the expected improvement in estimation performance
between smoothing and filtering has been realized,
particularly in terms of estimating the model index, $\auxvar_t$.

\section{Case study: log-polynomial systems}\label{sec:logpolynomial}
In this section, variational filtering is studied when the problem is log-polynomial of order, say $\nu$.
The model is then given by
\begin{subequations}
\begin{align}
\log \UnnormalizedDensity_0^0(\statevar_0) &=
\log \gaussian(\statevar_0; \StateMean_0, \StateCov_0)  + \sum_{m=0}^\nu \ell_0^{(m)} \statevar_0^m\\
\log \UnnormalizedDensity_{t\mid t-1}^t(\statevar_t \mid \statevar_{t-1}) &=
\log \gaussian(\statevar_t; \StateCondMean_{t\mid t-1}(\statevar_{t-1}), \StateCondCov_{t\mid t-1})  + \sum_{m=0}^\nu \ell_t^{(m)} \statevar_t^m
\end{align}
\end{subequations}
and the state is thus a scalar.
Note that by defining $p_m(\statevar) = \statevar^m$ the likelihoods can be written as inner products according to
\begin{equation}\label{eq:log-polynomial-likelihood}
\log \ObsDensity_{t\mid t}(\statevar_t) = \sum_{m=0}^\nu \ell_t^{(m)} \statevar_t^m = \langle \ell_t, p(\statevar_t)\rangle.
\end{equation}

The aim of this section is to establish the forward value functional recursion when $\VariationalFamily_{0:T}$ is Gauss--Markov
and thus the following theorem.
\begin{theorem}\label{thm:log-polynomial:forward-value-functional}
Let the variational family $\VariationalFamily_{0:T}$ be Gauss--Markov then the forward value functional associated with a log-polynomial system of order $\nu \geq 2$ is given by
\begin{equation}
\log \ForwardRepresenter_t(\statevar_t) = \sum_{m=0}^\nu \lambda_t^{(m)} \statevar_t^m = \langle \lambda_t, p(\statevar_t) \rangle.
\end{equation}
\end{theorem}

\Cref{thm:log-polynomial:forward-value-functional} readily gives the filter update, namely
\begin{equation}
\begin{split}
\int \log \Big\{ \frac{\ForwardRepresenter_t(\statevar_t)}{\VariationalDensity_t(\statevar_t)} \Big\} \VariationalDensity_t(\statevar_t) \dif \statevar_t
&= \int \langle \lambda_t, p(\statevar_t) \rangle \VariationalDensity_t(\statevar_t) \dif \statevar_t - \int \log\{\VariationalDensity_t(\statevar_t)\} \VariationalDensity_t(\statevar_t) \dif \statevar_t.
\end{split}
\end{equation}
Applying \cref{lem:gaussian-non-central-moments} from \cref{app:log-polynomial} to the first term then gives the following result.

\begin{proposition}[Variational filter update]
Let $\VariationalDensity_t$ ge Gaussian with mean and covariance $\StateMean_t$ and $\StateCov_t$, respectively, then
\begin{equation}
\begin{split}
\int \log \Big\{ \frac{\ForwardRepresenter_t(\statevar_t)}{\VariationalDensity_t(\statevar_t)} \Big\} \VariationalDensity_t(\statevar_t) \dif \statevar_t
&= \langle \lambda_t, A_1(\StateCov_t) p(\StateMean_t) \rangle + \frac{1}{2}(\log\{2\pi \Sigma_t\} + 1),
\end{split}
\end{equation}
where $A_1$ is defined in \cref{lem:gaussian-non-central-moments}.
Hence the variational filter mean and covariance satisfies
\begin{equation}
\StateMean_t^{0:t},\, \StateCov_t^{0:t} = \argmax[\StateMean_t, \StateCov_t] \big\{ \langle \lambda_t, A_1(\StateCov_t) p(\StateMean_t) \rangle + \frac{1}{2}(\log\{2\pi \Sigma_t\} + 1) \big\}.
\end{equation}
\end{proposition}

It now remains to establish \cref{thm:log-polynomial:forward-value-functional} and a recursion for the value functional.
Note that \cref{thm:log-polynomial:forward-value-functional}  hold at $t = 0$ by the following identification
\begin{subequations}
\begin{align}
\lambda_0^{(0)} &= -\frac{1}{2}\log \{2\pi \StateCov_0\} - \frac{\StateMean_0^2}{2\StateCov_0} + \ell_0^{(0)} \\
\lambda_0^{(1)} &= \frac{\StateMean_0}{\StateCov_0} + \ell_0^{(1)} \\
\lambda_0^{(2)} &= -\frac{1}{2\StateCov_0} + \ell_0^{(2)} \\
\lambda_0^{(m)} &= \ell_0^{(m)}, \quad 2 \leq m.
\end{align}
\end{subequations}
It then remains to establish the induction step.
The value functional recursion is
\begin{equation}
\begin{split}
\log \ForwardRepresenter_t(\statevar_t) &= \int \log \Big\{ \frac{\UnnormalizedDensity_{t\mid t-1}^t(\statevar_t \mid \statevar_{t-1})\ForwardRepresenter_{t-1}(\statevar_{t-1})}
{\VariationalDensity_{t-1\mid t}(\statevar_{t-1} \mid \statevar_t)} \Big\} \VariationalDensity_{t-1\mid t}(\statevar_{t-1} \mid \statevar_t) \dif \statevar_{t-1} \\
&= \log \ObsDensity_{t\mid t}(\statevar_t) +   \int \log \Big\{ \frac{\StateDensity_{t\mid t-1}(\statevar_t \mid \statevar_{t-1})}
{\VariationalDensity_{t-1\mid t}(\statevar_{t-1} \mid \statevar_t)} \Big\} \VariationalDensity_{t-1\mid t}(\statevar_{t-1} \mid \statevar_t) \dif \statevar_{t-1} \\
&\quad + \int \log \{ \ForwardRepresenter_{t-1}(\statevar_{t-1}) \} \VariationalDensity_{t-1\mid t}(\statevar_{t-1} \mid \statevar_t) \dif \statevar_{t-1} \\
&= \langle \ell_t, p(\statevar_t) \rangle + \langle c_t(\TransitionMatrix_{t-1, t}, \StateControl_{t-1, t}, \StateCondCov_{t-1\mid t}), p(\statevar_t) \rangle \\
&\quad + \int \log \{ \ForwardRepresenter_{t-1}(\statevar_{t-1}) \} \VariationalDensity_{t-1\mid t}(\statevar_{t-1} \mid \statevar_t) \dif \statevar_{t-1} \\
\end{split}
\end{equation}
where \eqref{eq:log-polynomial-likelihood} was used for the first term and $c_t$ is defiend by \cref{lem::gaussian-cross-entropy} in \cref{app:log-polynomial}.
Additionally, by \cref{lem:gaussian-non-central-moments} in \cref{app:log-polynomial}, there are matrices $A_1(\StateCondCov_{t-1\mid t})$ and $A_2(\TransitionMatrix_{t-1, t}, \StateControl_{t-1,t})$ such that
\begin{equation}
\begin{split}
\int \langle \lambda_{t-1}, p(\statevar_{t-1}) \rangle   \VariationalDensity_{t-1\mid t}(\statevar_{t-1} \mid \statevar_t) \dif \statevar_{t-1}
&= \langle \lambda_{t-1}, A_1(\StateCondCov_{t-1\mid t}) p(\StateCondMean_{t-1\mid t}(\statevar_t)) \rangle \\
&=  \langle \lambda_{t-1}, A_1(\StateCondCov_{t-1\mid t}) A_2(\TransitionMatrix_{t-1, t}, \StateControl_{t-1,t}) p(\statevar_t) \rangle
\end{split}
\end{equation}
\Cref{thm:log-polynomial:forward-value-functional} has then been established along with the following value functional recursion.

\begin{proposition}[Forward value functional update]
Assume \cref{thm:log-polynomial:forward-value-functional} hold at $t-1$ then it also hold at $t$.
Additionally the parameter $\lambda_t$ can be obtained from $\lambda_{t-1}$ according to
\begin{equation}
\lambda_t =A_2^*(\TransitionMatrix_{t-1, t}, \StateControl_{t-1,t})  A_1^*(\StateCondCov_{t-1\mid t}) \lambda_{t-1}
+ c_t(\TransitionMatrix_{t-1, t}, \StateControl_{t-1, t}, \StateCondCov_{t-1\mid t})
+ \ell_t
\end{equation}
where the matrices $A_1$ and $A_2$ are defined in \cref{lem:gaussian-non-central-moments}, and $c_t$ is defined in \cref{lem::gaussian-cross-entropy}.
\end{proposition}

\subsection{Simulation study: The cubic sensor problem}
Let  $\StateCondMean_{t\mid t-1}(\statevar) = \phi_0 \statevar + (1 - \phi_0)\mu_0$ and $\ObsCondMean_{t\mid t}(\statevar) = \beta \statevar^3$ and consider
the cubic sensor system given by \citep{Katayama2013}
\begin{subequations}
\begin{align}
\statevar_0 &\sim \gaussian(\StateMean_0, \StateCov_0) \\
\statevar_t \mid \statevar_{t-1} &\sim \gaussian(\statevar_t; \StateCondMean_{t\mid t-1}(\statevar_{t-1}), \StateCondCov_{t\mid t-1}) \\
\obsvar_t \mid \statevar_t &\sim \gaussian(\obsvar_t; \ObsCondMean_{t\mid t}(\statevar_t), \ObsCondCov).
\end{align}
\end{subequations}
It is clearly a log-polynomial system of order $6$ with the likelihoods given by
\begin{equation}\label{eq:cubic-sensor-likelihood}
\begin{split}
\log \ObsDensity_{t\mid t}(\statevar_t) &= - \frac{1}{2} \Big( \log\{2\pi \ObsCondCov \} + \frac{\obsvar_t^2}{2\ObsCondCov}  \Big) +  \frac{\obsvar_t \beta}{\ObsCondCov} \statevar_t^3
- \frac{\beta^2}{2\ObsCondCov} \statevar_t^6 \\
&= \sum_{m=0}^6 \ell_t^{(m)} \statevar_t^m.
\end{split}
\end{equation}
It is interesting to note that this sytsem has two regimes. Namely, $(i)$ $\abs{\statevar_t} < 1$ where the signal-to-noise ratio is small and $(ii)$ $\abs{\statevar_t} \geq 1$
where the signal-to-noise-ratio is high.
In particular, if $\mu_0 = 0$ then $\statevar_t$ is a zero mean stationary process, which together with \eqref{eq:cubic-sensor-likelihood} means that
the posterior $\StateDensity_{0:T}^{0:T}$ has a mode at the zero-path, $\statevar_t = 0, \quad t \geq 0$.

The parameters are set to $\Sigma_0 = 0.36$, $\phi_0 = 0.95$, $\StateCondCov_{t\mid t-1} = (1 - \abs{\phi_0}^2) \Sigma_0$, $\beta = 1$, $R =  1$,
and the mean parameter is varied according to $\mu_0 \in \{0, 0.5, 1.5\}$.
Thus when $\mu_0 = 0$, the state has a 10\% chance of being greater than unity in magnitude.
The system is for $T = 2^{13}+1$ time points for each configuration.

The sub-optimal variational filter of \cref{def:sub-optimal-variational-filter-1} is implemented and the optimization is carried out with natural gradient descent
using a step-size of $\num{1e-1}$ and 20 iterations per filter step.
The sub-optimal filter is compared to an iterated extended Kalman filter \citep{Bell1993}, which also uses 20 iterations.
The two alternatives are referred to as VF(20) and EKF(20), respectively.

\begin{figure}
\centering
\includegraphics{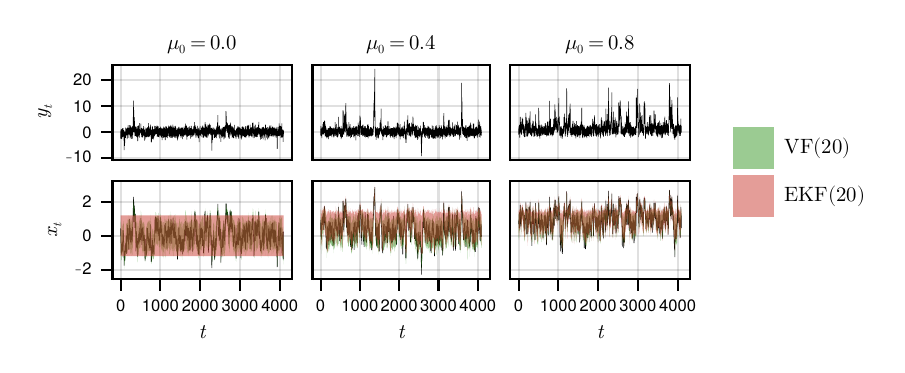}
\caption{
Simulation results for the cubic sensor experiment.
The rop row shows the realization of the observation sequence for each selection of $\mu_0$ and the bottom row shows the corresponding realization of the state sequence.
The filter estimates are illustrated by a two standard deviation band around their estimates.
}\label{fig:logpolynomial}
\end{figure}

The results of the three simulations are shown in \cref{fig:logpolynomial}.
It can be seen that EKF(20) fails to track the state completely in the case when the stationary mean of the state is zero (it gets stuck at the mode at zero).
However, as the stationary mean moves away from zero VF(20) and EKF(20) become increasingly similar in performance.
This is also corroborated by the sequence of root mean-square errors, which is shown in \cref{tab:logpolynomial:rmse}.
The sum of $\chi^2$-statistics is shown in \cref{tab:logpolynomial:chi2} from which a similar conclusion can be drawn
but also that both filters are quite severely overconfident on average.
In fact, the overconfidence gets worse the more time the state stays in the high signal-to-noise ratio regime.

\begin{table}[t!]
\centering
\caption{
Root mean-square errors for VF(20) and EKF(20) in the cubic sensor problem. \\
It can be seen that VF(20) handles the cases when the state is seldom in the high signal-to-noise ratio regime better than EKF(20). \\
}
\begin{tabular}{l|c|c|c}
\hline
$\mu_0$ & 0 & 0.4 & 0.8 \\ \hline
VF(20)  & 0.411192 &  0.380229 &  0.33623 \\ \hline
EKF(20) & 0.614548 &  0.527413 &  0.410902 \\ \hline
\end{tabular}\label{tab:logpolynomial:rmse}
\end{table}

\begin{table}[t!]
\centering
\caption{
Sum of $\chi^2$-statistics for VF(20) and EKF(20) in the cubic sensor problem.
If a Gaussian filter is appropriate this sum should with high probability be $T = 2^{13}+1 = 8193$\\
}
\begin{tabular}{l|c|c|c}
\hline
$\mu_0$ & 0 & 0.4 & 0.8 \\ \hline
VF(20)  & 11466.5 &  13863.3 &  20122.9 \\ \hline
EKF(20) & 8595.12 &  27356.0 &  19681.2 \\ \hline
\end{tabular}\label{tab:logpolynomial:chi2}
\end{table}

\section{Conclusion}
In this article, a complete set of variational analogue to Bayesian state estimation theory was established.
In particular, the value functional associated with the backward method, $\BackwardRepresenter_t$, was shown to be a lower bound on the state likelihood $\ObsDensity_{t+1:T\mid t}$,
and the value functional associated with the forward method, $\ForwardRepresenter_t$, was shown to be a lower bound on the unnormalized filtering density $\UnnormalizedDensity_t^{0:t}$.
Like the Bayesian state estimation recursions, the variational state estimation recursion are intractable in general.
However, it has been demonstrated that they are tractable \emph{more often} than their Bayesian counterparts.
Additionally, a formal notion of sub-optimal variational filtering has been put forth, which ensures a time-complexity of $O(T)$.
Notably this notion of variational filtering differs from that of \citet{Campbell2021}.
The methodology has been empirically demonstrated to be sound in a jump Gauss--Markov system with a factored approximation and in a Gauss--Markov model with log-polynomial likelihoods under a Gauss--Markov constraint.
There are several issues and possibilities that can be pursued in future work.
From a methodological point of view, the two most pressing issues appear to be:

\begin{itemize}
\item \Cref{prop:iterative-backward-smoother,prop:iterative-forward-smoother} need to compute integrals with respect to all time-marginals of the class $\VariationalFamily_{0:T}$.
Unfortunately, these time-marginals are analytically tractable only in a few cases such as when $\VariationalFamily_{0:T}$ is a Gauss--Markov class.
\item The value functional recursions are intractable. \Cref{app:forward-value-functional-approximation-1} is a great start as demonstrated by its implicit use by \citet{Courts2021}.
However, approximation methods where the error can be reduced or even controlled could prove significant in improving state estimation performance.
\end{itemize}

A possibility of resolving the first issue is to approximate intergrals with respect to the time-marginals with Monte Carlo averages.
As the value functional recursions of \cref{thm:variational-backward-hjb,thm:variational-forward-hjb} are given in the \emph{weak} sense,
an immediate approach would be to resort to some Galerkin style approximation with an appropriate selection of subspace.
However, these possibilities will have to explored in future work.

\section*{Acknowledgements}
FT was partially supported by the Wallenberg AI, Autonomous Systems and Software Program (WASP) funded by the Knut and Alice Wallenberg Foundation.

\appendix

\section{More details on jump Gauss--Markov processes}\label{app:vjgm}

In this section, some additonal details are provided for variational state estimation in jump Gauss--Markov systems.
More specifically formulae are provided for conveniently organizing the implementation.
The proofs of these results reduce to rote algebra and are therefore omitted.

\begin{lemma}
Let $g^\star(\statevar) = \gaussian(\StateMean^\star, \StateCov^*)$ and $g(\statevar) = \gaussian(\StateMean, \StateCov)$
then the negative relative entropy between of $g^\star$ with respect to $g$ is given by
\begin{equation}\label{eq:normal-relative-entropy}
\begin{split}
&\int \log \Big\{\frac{g^\star(\statevar)}{g(\statevar)}\Big\} g(\statevar)\dif \statevar \\
&= -\frac{1}{2} \Big[ \tr( \{\StateCov^\star\}^{-1} \StateCov )  - \tr\{\Id\} +
(\StateMean^\star - \StateMean)^* \{\StateCov^\star\}^{-1}  (\StateMean^\star - \StateMean)
+ \log \abs[0]{\StateCov^\star }  - \log \abs[0]{\StateCov} \Big].
\end{split}
\end{equation}
\end{lemma}

\begin{lemma}[Average of log-Gaussian densities]\label{lem:average-log-gaussian}
Let $g(\statevar \mid \auxvar) = \gaussian(\statevar; \StateMean(\auxvar), \StateCov(\auxvar))$
and $f(z)$ a categorical distribution.
Then there is a $\bar{g}(\statevar) = \gaussian(\bar{\StateMean}, \bar{\StateCov})$ and $\zeta$ such that
\begin{equation}
\log \{\zeta \bar{g}(\statevar)\} = \sum_z \log g(\statevar\mid \auxvar) f(\auxvar).
\end{equation}
Furthermore, the parameters $\bar{\StateMean}, \bar{\StateCov},$ and $\zeta$ are given by
\begin{subequations}
\begin{align}
\bar{\StateCov}^{-1} &= \sum_\auxvar \StateCov^{-1}(\auxvar) f(\auxvar) \\
\bar{\StateCov}^{-1} \bar{\StateMean} &=  \sum_\auxvar \StateCov^{-1}(\auxvar) \StateMean(\auxvar) f(\auxvar) \\
\zeta &= \sum_z \log g(\bar{\StateMean} \mid \auxvar) f(\auxvar) + \frac{1}{2} \log \abs[0]{2\pi\bar{\StateCov}}.
\end{align}
\end{subequations}
\end{lemma}

\begin{lemma}[Negative relative entropy of Gaussian conditionals]\label{lem:relative-entropy-likelihood}
Let $\kappa_1(\statevar^\prime \mid \statevar) = \gaussian(\statevar^\prime; \StateCondMean_1(\statevar), \StateCondCov_1)$
and $\kappa_2(\statevar^\prime \mid \statevar) = \gaussian(\statevar^\prime; \StateCondMean_2(\statevar), \StateCondCov_2)$
with $\StateCondMean_i$ an affine function with slope and intercept given by $\TransitionMatrix_i$ and $\StateControl_i$, respectively for $i=1,2$.
Then the negative relative entropy of $\kappa_2$ with respect to $\kappa_1$ is given by
\begin{equation}
\int \log \Big\{ \frac{\kappa_2(\statevar^\prime \mid \statevar) }{\kappa_1(\statevar^\prime \mid \statevar)} \Big\} \kappa_1(\statevar^\prime \mid \statevar) \dif \statevar^\prime  =
c \mathcal{N}(\obsvar; \ObsMatrix \statevar, \StateCondCov_2)
\end{equation}
where the parameters $c$, $y$, and $C$ are given by
\begin{subequations}
\begin{align}
\log c &= - \frac{1}{2} \Big[ \tr\{\StateCondCov_2^{-1} \StateCondCov_1 \} - \tr\{\Id\}(1 + \log2\pi) - \log \abs{\StateCondCov_1} \Big] \\
\obsvar &= \StateControl_2 - \StateControl_1 \\
\ObsMatrix &= \TransitionMatrix_1 - \TransitionMatrix_2.
\end{align}
\end{subequations}
\end{lemma}

\section{More details on log-polynomial systems}\label{app:log-polynomial}

In this section additional details are givne for log-polynomial systems.
Some results rely on the classic formula for central moments of univariate Gaussian distributions,
which is stated in \cref{lem:gaussian-moments}.

\begin{lemma}[Gaussian central moments]\label{lem:gaussian-moments}
Let $\VariationalDensity(\statevar) = \gaussian(\statevar; \mu, \Sigma)$ then the central moments are given by
\begin{equation}
\int (\statevar - \mu)^p \VariationalDensity(\statevar) \dif \statevar =
\begin{cases}
0, \quad p \text{ odd}\\
\Sigma^{p/2} (p-1)!!, \quad p \text{ even}
\end{cases},
\end{equation}
where $!!$ denotes the double factorial.
\end{lemma}

\begin{lemma}\label{lem:gaussian-non-central-moments}
Let $\kappa(\statevar \mid \auxvar) = \gaussian(\statevar; \StateCondMean(\auxvar), \StateCondCov)$ with conditional mean function
$\StateCondMean(\statevar) = \TransitionMatrix \statevar + \StateControl$ and define the matrix $A_1(Q)$ by
\begin{subequations}
\begin{align}
\{A_1(\StateCondCov)\}_{n, m} &=
\begin{cases}
{n \choose m} \StateCondCov^{\frac{n-m}{2}}(n - m - 1)!!, \quad n - m \text{ even} \\
0, \quad n -m \text{ odd or } m > n
\end{cases} \\
\{A_2(\TransitionMatrix, \StateControl)\}_{n, m} &= {n \choose m} \TransitionMatrix^m \StateControl^{n-m}.
\end{align}
\end{subequations}
Then the following hold
\begin{equation}
\int p(\statevar) \kappa(\statevar \mid \auxvar) \dif \statevar = A_1(\StateCondCov) p(\StateCondMean(\auxvar)) = A_1(\StateCondCov) A_2(\TransitionMatrix, \StateControl) p(\auxvar)
\end{equation}
\end{lemma}

\begin{lemma}\label{lem::gaussian-cross-entropy}
Let $\kappa_1(\statevar \mid \auxvar) = \gaussian(\statevar; \StateCondMean_1(\auxvar), \StateCondCov_1)$
and $\kappa_2(\auxvar \mid \statevar) = \gaussian(\auxvar; \StateCondMean_2(\statevar), \StateCondCov_2)$
with $\StateCondMean_i$ being an affine map with slope and intercept given by $\TransitionMatrix_i$ and $\StateControl_i$, respectively, for $i = 1, 2$.
Then the negative cross-entropy of $\kappa_2$ relative to $\kappa_1$ is given by
\begin{equation*}
\begin{split}
\int \log\Big\{ \frac{\kappa_2(\auxvar\mid \statevar)}{\kappa_1(\statevar\mid \auxvar)} \Big\} \kappa_1(\statevar\mid \auxvar) \dif \statevar  &=
-\frac{1}{2} \Big\{ \log\abs{\StateCondCov_2} - \log \abs{\StateCondCov_1} +  \tr(\StateCondCov_2^{-1}\TransitionMatrix_2 \StateCondCov_1 \TransitionMatrix_2^*) - \tr(\StateCondCov_1^{-1} \StateCondCov_1) \Big\} \\
&\quad - \frac{1}{2} \big \langle \auxvar - \StateCondMean_2\circ\StateCondMean_1(\auxvar), \StateCondCov_2^{-1}, (\auxvar - \StateCondMean_2\circ\StateCondMean_1(\auxvar))  \big \rangle .
\end{split}
\end{equation*}
Additionally, in the scalar case this evaluates to a polynomial of degree two with coefficents given by
\begin{subequations}
\begin{align}
c^{(0)}(\TransitionMatrix_1, \StateControl_1, \StateCondCov_1) &=
-\frac{1}{2} \Big\{ \log \StateCondCov_2 - \log \StateCondCov_1 +   \frac{ \abs{\TransitionMatrix_2}^2 \StateCondCov_1}{\StateCondCov_2} - 1 + \frac{\abs{\StateCondMean_2(\StateControl_1)}^2}{\StateCondCov_2} \Big\} \\
c^{(1)} (\TransitionMatrix_1, \StateControl_1, \StateCondCov_1) &= \frac{(1 - \TransitionMatrix_2\TransitionMatrix_1) \StateCondMean_2(\StateControl_1)}{\StateCondCov_2} \\
c^{(2)} (\TransitionMatrix_1, \StateControl_1, \StateCondCov_1) &= -\frac{\abs{1 - \TransitionMatrix_2\TransitionMatrix_1}^2}{2\StateCondCov_2}
\end{align}
\end{subequations}
\end{lemma}

\bibliographystyle{apalike}
\bibliography{refs}

@article{Abdulsamad2025,
  title={Proximal Approximate Inference in State-Space Models},
  author={Abdulsamad, Hany and Garc{\'\i}a-Fern{\'a}ndez, {\'A}ngel F and S{\"a}rkk{\"a}, Simo},
  journal={arXiv preprint arXiv:2511.15409},
  year={2025}
}

@article{Agamennoni2012,
  title={Approximate inference in state-space models with heavy-tailed noise},
  author={Agamennoni, Gabriel and Nieto, Juan I and Nebot, Eduardo M},
  journal={IEEE Transactions on Signal Processing},
  volume={60},
  number={10},
  pages={5024--5037},
  year={2012},
  publisher={IEEE}
}

@article{Arulampalam2002,
  title={A tutorial on particle filters for online nonlinear/non-{G}aussian {B}ayesian tracking},
  author={Arulampalam, M Sanjeev and Maskell, Simon and Gordon, Neil and Clapp, Tim},
  journal={IEEE Transactions on Signal Processing},
  volume={50},
  number={2},
  pages={174--188},
  year={2002},
  publisher={Ieee}
}

@article{Askar1981,
  title={A recursive algorithm for the {B}ayes solution of the smoothing problem},
  author={Askar, M. and Derin, H.},
  journal={IEEE Transactions on Automatic Control},
  volume={26},
  number={2},
  pages={558--561},
  year={1981},
  publisher={IEEE}
}

@article{Balenzuela2022,
  title={A new smoothing algorithm for jump {M}arkov linear systems},
  author={Balenzuela, Mark Peter and Wills, Adrian G and Renton, Christopher and Ninness, Brett},
  journal={Automatica},
  volume={140},
  pages={110218},
  year={2022},
  publisher={Elsevier}
}

@book{Barfoot2017,
  title={State Estimation for Robotics},
  author={Barfoot, T. D.},
  year={2017},
  publisher={Cambridge University Press}
}

@book{BarShalom2004,
  title={Estimation with Applications to Tracking and Navigation: Theory, Algorithms and Software},
  author={Bar--Shalom, Yaakov and Li, X Rong and Kirubarajan, Thiagalingam},
  year={2004},
  publisher={John Wiley \& Sons}
}

@book{Beal2003,
  title={Variational algorithms for approximate Bayesian inference},
  author={Beal, Matthew James},
  year={2003},
  publisher={University of London, University College London (United Kingdom)}
}

@article{Bell1993,
  title={The iterated {K}alman filter update as a {G}auss-{N}ewton method},
  author={Bell, Bradley M and Cathey, Frederick W},
  journal={IEEE Transactions on Automatic Control},
  volume={38},
  number={2},
  pages={294--297},
  year={1993},
  publisher={IEEE}
}

@article{Bell1994,
  title={The iterated {K}alman smoother as a {G}auss--{N}ewton method},
  author={Bell, Bradley M},
  journal={SIAM Journal on Optimization},
  volume={4},
  number={3},
  pages={626--636},
  year={1994},
  publisher={SIAM}
}

@article{Bellman1966,
  title={Dynamic programming},
  author={Bellman, Richard},
  journal={Science},
  volume={153},
  number={3731},
  pages={34--37},
  year={1966},
  publisher={American Association for the Advancement of Science}
}

@book{Bishop2006,
  title={Pattern Recognition and Machine Learning},
  author={Bishop, Christopher M and Nasrabadi, Nasser M},
  volume={4},
  number={4},
  year={2006},
  publisher={Springer}
}

@article{Blei2017,
  title={Variational inference: A review for statisticians},
  author={Blei, David M and Kucukelbir, Alp and McAuliffe, Jon D},
  journal={Journal of the American statistical Association},
  volume={112},
  number={518},
  pages={859--877},
  year={2017},
  publisher={Taylor \& Francis}
}

@article{Bresler1986,
  title={Two-filter formulae for discrete-time non-linear {B}ayesian smoothing},
  author={Bresler, Y.},
  journal={International Journal of Control},
  volume={43},
  number={2},
  pages={629--641},
  year={1986},
  publisher={Taylor \& Francis}
}

@article{Briers2010,
  title={Smoothing algorithms for state--space models},
  author={Briers, M. and Doucet, A. and Maskell, S.},
  journal={Annals of the Institute of Statistical Mathematics},
  volume={62},
  number={1},
  pages={61--89},
  year={2010},
  publisher={Springer}
}

@article{Campbell2021,
  title={Online variational filtering and parameter learning},
  author={Campbell, Andrew and Shi, Yuyang and Rainforth, Thomas and Doucet, Arnaud},
  journal={Advances in Neural Information Processing Systems},
  volume={34},
  pages={18633--18645},
  year={2021}
}

@book{Cappe2005,
  title={Inference in Hidden Markov Models},
  author={Capp{\'e}, O. and Moulines, E. and Ryd\'{e}n, T.},
  year={2005},
  publisher={Springer}
}

@article{Cappe2007,
  title={An overview of existing methods and recent advances in sequential {M}onte {C}arlo},
  author={Capp{\'e}, Olivier and Godsill, Simon J and Moulines, Eric},
  journal={Proceedings of the IEEE},
  volume={95},
  number={5},
  pages={899--924},
  year={2007},
  publisher={IEEE}
}

@article{Carrassi2018,
  title={Data assimilation in the geosciences: An overview of methods, issues, and perspectives},
  author={Carrassi, Alberto and Bocquet, Marc and Bertino, Laurent and Evensen, Geir},
  journal={Wiley Interdisciplinary Reviews: Climate Change},
  volume={9},
  number={5},
  pages={e535},
  year={2018},
  publisher={Wiley Online Library}
}

@article{Chagneux2024,
  title={Additive smoothing error in backward variational inference for general state-space models},
  author={Chagneux, Mathis and Gassiat, {\'E}lisabeth and Gloaguen, Pierre and Le Corff, Sylvain},
  journal={Journal of Machine Learning Research},
  volume={25},
  number={28},
  pages={1--33},
  year={2024}
}

@inproceedings{Corenflos2023,
  title={Variational {G}aussian filtering via {W}asserstein gradient flows},
  author={Corenflos, Adrien and Abdulsamad, Hany},
  booktitle={2023 31st European Signal Processing Conference (EUSIPCO)},
  pages={1838--1842},
  year={2023},
  organization={IEEE}
}

@article{Courts2021,
  title={Gaussian variational state estimation for nonlinear state-space models},
  author={Courts, Jarrad and Wills, Adrian G and Sch{\"o}n, Thomas B},
  journal={IEEE Transactions on Signal Processing},
  volume={69},
  pages={5979--5993},
  year={2021},
  publisher={IEEE}
}

@article{Crisan2002,
  title={A survey of convergence results on particle filtering methods for practitioners},
  author={Crisan, D. and Doucet, A.},
  journal={IEEE Transactions on Signal Processing},
  volume={50},
  number={3},
  pages={736--746},
  year={2002},
  publisher={IEEE}
}

@article{Darling2017,
  title={Minimization of the {K}ullback--{L}eibler divergence for nonlinear estimation},
  author={Darling, Jacob E and DeMars, Kyle J},
  journal={Journal of Guidance, Control, and Dynamics},
  volume={40},
  number={7},
  pages={1739--1748},
  year={2017},
  publisher={American Institute of Aeronautics and Astronautics}
}

@article{Djuric2003,
  title={Particle filtering},
  author={Djuric, Petar M and Kotecha, Jayesh H and Zhang, Jianqui and Huang, Yufei and Ghirmai, Tadesse and Bugallo, M{\'o}nica F and Miguez, Joaquin},
  journal={IEEE Signal Processing Magazine},
  volume={20},
  number={5},
  pages={19--38},
  year={2003},
  publisher={IEEE}
}

@book{Durbin2012,
  title={Time Series Analysis by State Space Methods},
  author={Durbin, James and Koopman, Siem Jan},
  year={2012},
  publisher={Oxford University Press (UK)}
}

@book{Evensen2009,
  title={Data assimilation: the ensemble {K}alman filter},
  author={Evensen, Geir},
  year={2009},
  publisher={Springer}
}

@article{Galka2004,
  title={A solution to the dynamical inverse problem of {EEG} generation using spatiotemporal {K}alman filtering},
  author={Galka, Andreas and Yamashita, Okito and Ozaki, Tohru and Biscay, Rolando and Vald{\'e}s-Sosa, Pedro},
  journal={NeuroImage},
  volume={23},
  number={2},
  pages={435--453},
  year={2004},
  publisher={Elsevier}
}

@article{Garcia2015,
  title={Posterior linearization filter: Principles and implementation using sigma points},
  author={Garc{\'\i}a-Fern{\'a}ndez, {\'A}ngel F and Svensson, Lennart and Morelande, Mark R and S{\"a}rkk{\"a}, Simo},
  journal={IEEE Transactions on Signal Processing},
  volume={63},
  number={20},
  pages={5561--5573},
  year={2015},
  publisher={IEEE}
}

@article{Garcia2016,
  title={Iterated posterior linearization smoother},
  author={Garc{\'\i}a-Fern{\'a}ndez, {\'A}ngel F and Svensson, Lennart and S{\"a}rkk{\"a}, Simo},
  journal={IEEE Transactions on Automatic Control},
  volume={62},
  number={4},
  pages={2056--2063},
  year={2016},
  publisher={IEEE}
}

@article{Garcia2019a,
  title={Gaussian target tracking with direction-of-arrival von {M}ises--{F}isher measurements},
  author={Garcia-Fernandez, Angel F and Tronarp, Filip and S{\"a}rkk{\"a}, Simo},
  journal={IEEE Transactions on Signal Processing},
  volume={67},
  number={11},
  pages={2960--2972},
  year={2019},
  publisher={IEEE}
}

@inproceedings{Gordon1993,
  title={Novel approach to nonlinear/non-{G}aussian {B}ayesian state estimation},
  author={Gordon, Neil J and Salmond, David J and Smith, Adrian FM},
  booktitle={IEE proceedings F (radar and signal processing)},
  volume={140},
  pages={107--113},
  year={1993},
  organization={IET}
}

@article{Guarniero2017,
  title={The iterated auxiliary particle filter},
  author={Guarniero, P. and Johansen, A. M. and Lee, A.},
  journal={Journal of the American Statistical Association},
  volume={112},
  number={520},
  pages={1636--1647},
  year={2017},
  publisher={Taylor \& Francis}
}

@article{Gultekin2017,
  title={Nonlinear {K}alman filtering with divergence minimization},
  author={Gultekin, San and Paisley, John},
  journal={IEEE Transactions on Signal Processing},
  volume={65},
  number={23},
  pages={6319--6331},
  year={2017},
  publisher={IEEE}
}

@article{Helmick1995,
  title={Fixed-interval smoothing for {M}arkovian switching systems},
  author={Helmick, Ronald E and Blair, W Dale and Hoffman, Scott A},
  journal={IEEE Transactions on Information Theory},
  volume={41},
  number={6},
  pages={1845--1855},
  year={1995},
  publisher={IEEE}
}

@article{Ho1964,
  title={A {B}ayesian approach to problems in stochastic estimation and control},
  author={Ho, YC and Lee, RCKA},
  journal={IEEE Transactions on Automatic Control},
  volume={9},
  number={4},
  pages={333--339},
  year={1964},
  publisher={IEEE}
}

@book{Hyndman2008,
  title={Forecasting with Exponential Smoothing: The State Space Approach},
  author={Hyndman, Rob and Koehler, Anne B and Ord, J Keith and Snyder, Ralph D},
  year={2008},
  publisher={Springer Science \& Business Media}
}

@article{Julier2000,
  title={A new method for the nonlinear transformation of means and covariances in filters and estimators},
  author={Julier, Simon and Uhlmann, Jeffrey and Durrant-Whyte, Hugh F},
  journal={IEEE Transactions on Automatic Control},
  volume={45},
  number={3},
  pages={477--482},
  year={2000},
  publisher={IEEE}
}

@book{Kailath2000,
  title={Linear estimation},
  author={Kailath, Thomas and Sayed, Ali H and Hassibi, Babak},
  year={2000},
  publisher={Prentice Hall}
}

@article{Kantas2009,
  title={An overview of sequential  {M}onte {C}arlo methods for parameter estimation in general state-space models},
  author={Kantas, N. and Doucet, A. and Singh, S. S. and Maciejowski, J. M.},
  journal={In Proceedings IFAC Symposium on System Identification (SYSID).},
  volume={42},
  number={10},
  pages={774--785},
  year={2009},
  publisher={Elsevier}
}

@inproceedings{Katayama2013,
  title={Equivalent linearization {K}alman filter with application to cubic sensor problems},
  author={Katayama, Tohru},
  booktitle={2013 European Control Conference (ECC)},
  pages={1633--1638},
  year={2013},
  organization={IEEE}
}

@inproceedings{Kim2020,
  title={Variational inference for sequential data with future likelihood estimates},
  author={Kim, Geon-Hyeong and Jang, Youngsoo and Yang, Hongseok and Kim, Kee-Eung},
  booktitle={International Conference on Machine Learning},
  pages={5296--5305},
  year={2020},
  organization={PMLR}
}

@article{Kitagawa1987,
  title={Non-{G}aussian state—space modeling of nonstationary time series},
  author={Kitagawa, Genshiro},
  journal={Journal of the American Statistical Association},
  volume={82},
  number={400},
  pages={1032--1041},
  year={1987},
  publisher={Taylor \& Francis}
}

@article{Larson1966,
  title={A dynamic programming approach to trajectory estimation},
  author={Larson, R. and Peschon, J.},
  journal={IEEE Transactions on Automatic Control},
  volume={11},
  number={3},
  pages={537--540},
  year={1966},
  publisher={IEEE}
}

@article{Lefebvre2002,
  title={Comment on" a new method for the nonlinear transformation of means and covariances in filters and estimators"[with authors' reply]},
  author={Lefebvre, Tine and Bruyninckx, Herman and De Schuller, Joris},
  journal={IEEE Transactions on Automatic Control},
  volume={47},
  number={8},
  pages={1406--1409},
  year={2002},
  publisher={IEEE}
}

@book{Lindstrom2018,
  title={Statistics for Finance},
  author={Lindstr{\"o}m, Erik and Madsen, Henrik and Nielsen, Jan Nygaard},
  year={2018},
  publisher={Chapman and Hall/CRC}
}

@article{Naesseth2019,
  title={Elements of sequential {M}onte {C}arlo},
  author={Naesseth, Christian A and Lindsten, Fredrik and Sch{\"o}n, Thomas B and others},
  journal={Foundations and Trends{\textregistered} in Machine Learning},
  volume={12},
  number={3},
  pages={307--392},
  year={2019},
  publisher={Now Publishers, Inc.}
}

@book{MarkleyCrassidis2014,
	author="Markley, F. L. and Crassidis, J. L.",
	title="Fundamentals of Spacecraft Attitude Determination",
	publisher="Springer",
	year = "2014"
}

@book{Maybeck1982,
	author= {Maybeck, P. S.},
	title= {Stochastic Models, Estimation and Control},
  volume={1-3},
	publisher={Academic Press},
	year = {1979,1982,1982}
}

@article{Pitt1999,
  title={Filtering via simulation: Auxiliary particle filters},
  author={Pitt, Michael K and Shephard, Neil},
  journal={Journal of the American statistical association},
  volume={94},
  number={446},
  pages={590--599},
  year={1999},
  publisher={Taylor \& Francis}
}

@book{Sarkka2023,
  title={Bayesian Filtering and Smoothing},
  author={S{\"a}rkk{\"a}, Simo and Svensson, Lennart},
  volume={2},
  year={2023},
  publisher={Cambridge university press}
}

@article{Smidl2008,
  title={Variational {B}ayesian filtering},
  author={Smidl, V{\'A}clav and Quinn, Anthony},
  journal={IEEE Transactions on Signal Processing},
  volume={56},
  number={10},
  pages={5020--5030},
  year={2008},
  publisher={IEEE}
}

@book{TittertonWeston2004,
	author="D. H. Titterton and J. L. Weston",
	title="Strapdown Inertial Navigation Technology",
	publisher="The Institute of Electrical Engineers",
	year = "2004"
}

@article{Tronarp2018,
  title={Iterative filtering and smoothing in nonlinear and non-{G}aussian systems using conditional moments},
  author={Tronarp, Filip and Garcia-Fernandez, Angel F and S{\"a}rkk{\"a}, Simo},
  journal={IEEE Signal Processing Letters},
  volume={25},
  number={3},
  pages={408--412},
  year={2018},
  publisher={IEEE}
}

@inproceedings{Tronarp2019b,
  title={Updates in {B}ayesian filtering by continuous projections on a manifold of densities},
  author={Tronarp, Filip and S{\"a}rkk{\"a}, Simo},
  booktitle={ICASSP 2019-2019 IEEE International Conference on Acoustics, Speech and Signal Processing (ICASSP)},
  pages={5032--5036},
  year={2019},
  organization={IEEE}
}

@inproceedings{Weber2015,
  title={Reinforced variational inference},
  author={Weber, Theophane and Heess, Nicolas and Eslami, Ali and Schulman, John and Wingate, David and Silver, David},
  booktitle={Advances in Neural Information Processing Systems (NIPS) Workshops},
  year={2015}
}

@article{Wu2006,
  title={A numerical-integration perspective on {G}aussian filters},
  author={Wu, Yuanxin and Hu, Dewen and Wu, Meiping and Hu, Xiaoping},
  journal={IEEE Transactions on Signal Processing},
  volume={54},
  number={8},
  pages={2910--2921},
  year={2006},
  publisher={IEEE}
}

@article{Zhao2020,
  title={Variational Online Learning of Neural Dynamics},
  author={Zhao, Yuan and Park, Il Memming},
  journal={Frontiers in Computational Neuroscience},
  volume={14},
  pages={71},
  year={2020},
  publisher={Frontiers Media SA}
}

\end{document}